%% file: pattern.tex
\documentclass[runningheads]{llncs}
\usepackage{graphicx}
\usepackage[T1]{fontenc}
\usepackage{cite}
\usepackage{amsmath, amssymb, amsfonts}
\usepackage{algorithmic}
\usepackage{graphicx}
\usepackage{textcomp}
\usepackage{pgfplots}
\pgfplotsset{compat=1.16}
\usepackage{xcolor}
\usepackage{diagbox}
\usepackage[section]{placeins}
\usepackage{mathrsfs}
\usepackage{subfigure}
\usepackage{physics}
\usepackage{tikz}
\usepackage{scalefnt}
\usepackage{qcircuit}
\usepackage{tabularx} 
\usepackage{enumerate}
\usepackage{multirow}
 \usepackage{pdfpages}
\usepackage{rotating}
\usepackage{caption}
\usepackage{hyperref}
\usepackage{threeparttable}

\usepackage{svg}
\svgsetup{
    inkscapepath=i/svg-inkscape/
}
\svgpath{{svg/}}

\usepackage[ruled, vlined]{algorithm2e}

\newcommand{\Cc}{\mathcal{C}}
\newcommand{\Pat}{\mathcal{R}}
\newcommand{\Ss}{\mathcal{S}}

\newcommand{\M}{\mathcal{M}}
\newcommand{\D}{\mathcal{D}}

\newcommand{\leaveout}[1]{}
\newcommand{\RNum}[1]{\uppercase\expandafter{\romannumeral #1\relax}}

\begin{document}
\begin{sloppypar}

\title{A Pattern Matching-Based Framework for Quantum Circuit Rewriting}
%
%
\author{Hui Jiang,  Diankang Li, Yuxin Deng,  Ming Xu}
\authorrunning{H. Jiang et al.}
%
\institute{Shanghai Key Laboratory of Trustworthy Computing, East China Normal University}

\maketitle              

\begin{abstract}The realization of quantum algorithms relies on specific quantum compilations according to the underlying quantum processors. 
However, there are various ways to physically implement qubits in different physical devices
and manipulate those qubits.
These differences lead to different communication methods and connection topologies, 
with each vendor implementing 
its own set of primitive gates.
Therefore, quantum circuits have to be rewritten or transformed in order to be transplanted from one platform to another.
We propose a pattern matching-based framework for rewriting quantum circuits, called QRewriting.
It takes advantage of a new representation of quantum circuits using symbol sequences.
Unlike the traditional way of using directed acyclic graphs,
the new representation allows us to easily identify  the patterns that appear non-consecutively but reducible.
Then, we convert the problem of pattern matching into that of finding distinct subsequences, 
and propose a polynomial-time dynamic programming-based pattern matching and replacement algorithm. 
We develop a rule library for basic optimizations and use it to rewrite the Arithmetic and Toffoli benchmarks from the $G_{IBM}$ gate set to the $G_{Sur}$ gate set.
Compared with the existing tool PaF, 
QRewriting obtains an improvement of reducing depths (resp. gate counts) by 29\% (resp. 14\%).

\end{abstract}

\section{Introduction}
\label{intro}
Quantum computing has attracted more and more interest in the last decades, 
since it provides the possibility to efficiently solve important problems
such as integer factorization~\cite{Sho94}, 
unstructured search~\cite{Gro96}, 
and solving linear equations~\cite{HHL09}. 

In recent years, with the popularity of quantum computing, many companies,
universities and institutes are actively working to develop prototypes
of quantum computer systems.
For example, in 2019, Google announced the realization of quantum supremacy,
the development of 53-qubit quantum processor ``Sycamore''~\cite{Arute19}.
In November 2021, 
IBM uneviled its new 127-qubit ``Eagle'' processor
whose scale makes it impossible for a classical computer to reliably simulate, 
and the increased qubit count allows users to explore problems at a new level of complexity~\cite{IBM127}.
In June 2022, Xanadu demonstrated a quantum computational advantage with a programmable photonic processor that realized Gaussian boson sampling on 216 sequeezed modes \cite{Mad22}.
These systems are referred to as Noisy Intermediate-Scale Quantum (NISQ) systems~\cite{Preskill2018} and have small qubit counts, 
restricted connectivity and high gate error rates. 
The coherence time of each physical qubit must be at least $1$--$10$ns, 
if the minimum physical gate fidelity of 99\% is to be achieved.
At present, the duration of physical quantum gate is $10$--$100$us and
only a limited set of quantum gates can be realized with relatively high fidelity on a quantum device~\cite{Kjaergaard2020}.
Each quantum processor may support a specific universal set of 1-qubit and 2-qubit gates, which are called primitive gates~\cite{Kjaergaard20}. 
Table~\ref{gateset} lists three gate sets: 
$G_{Com},\ G_{IBM}$ and $G_{Sur}$, where
$G_{Com}$ is a universal gate set~\cite{Nielsen2016},
 $G_{IBM}$ is implemented by the IBM QX5 quantum processor~\cite{Dumitrescu_2018},
and $G_{Sur}$ is used by the Surface-17 quantum processor~\cite{LaoSAA22}.

The realization of quantum algorithms relies on specific quantum compilations according to the underlying quantum processors.
However, there are various ways to physically implement qubits in different physical devices
and manipulate those qubits.
These differences lead to different communication methods and connection topologies, 
with each vendor implementing 
its own set of primitive gates.
Therefore, quantum circuits have to be transplanted or transplanted from one platform to another.
In addition, since the gate types supported by a quantum processor is limited,
quantum circuits may also be rewritten when some high-level gates are decomposed into low-level gates
before the quantum circuits can be executed on the quantum processor.

Converting a  quantum circuit supported by one gate set to a quantum circuit supported by another gate set with respect to some rules is called quantum circuit rewriting. Usually a rule is in the form $\Cc_p = \Cc_s$, where $\Cc_p$ is a fragment of a circuit whose behaviour is the same as that of the fragment $\Cc_s$. We call $\Cc_p$ a pattern circuit and $\Cc_s$ a substitution circuit. 
In this paper, we refer to the  circuit to be rewritten as the target circuit.
Motivated by the aforementioned requirements, our approach consists of two key steps:
one is to identify the desired patterns in the target circuit, 
the other is to replace them
with semantically equivalent substitution circuits.
For that purpose, we first introduce a new representation of quantum circuits using symbol sequences.
Unlike the traditional way of using directed acyclic graphs (DAGs), the new representation allows us to easily identify the patterns
that appear non-consecutively but reducible.
In the case that a pattern can be matched by several different rules,
we encounter a replacement conflict and need to resolve it with an appropriate policy.
We propose three policies for generating schedulers to cope with the replacement conflicts.
One policy is precise in the sense that it will consider all the replacement candidates of a conflict set.
In the worst case, its time complexity is exponential.
For a large-scale circuit, we need to make a trade-off between the quality of the generated circuit and the time it takes. 
Therefore, for large-scale circuits, we propose a greedy and a stochastic policy to handle the replacement conflicts.

\begin{table}[t]
    \caption{\label{gateset}Gate sets used in our evaluation. }\vspace{1mm}
 \renewcommand\arraystretch{1.2}
    \begin{tabularx}{\linewidth}{l|X}
            \hline
         $G_{Com}$~\cite{Nielsen2016} \quad\quad  &\quad H, X, Y, Z, S, S$^\dagger$ T, T$^\dagger$, {\rm R$_{\rm z}(\theta)$}, CX \\
         $G_{IBM}$~\cite{Dumitrescu_2018} &\quad   U1($\theta$), U2($\phi$, $\lambda$), U3($\theta$, $\phi$, $\lambda$), CX \\
         $G_{Sur}$~\cite{LaoSAA22}\quad \quad &\quad   X, Y, {\rm R$_{\rm x}(\theta)$}, {\rm R$_{\rm y}(\theta)$}, CZ\\
         \hline
            \end{tabularx}
\end{table}

The main contributions of this paper are listed below.
\begin{itemize}
    \item We introduce  a new representation of quantum circuits,
    which can easily identify the patterns that appear non-consecutively but remain reducible in the target circuits.
    \item We present a polynomial-time dynamic programming-based pattern matching and replacement algorithm.
    \item We propose three policies for generating schedulers to deal with replacement conflicts.
    \item We develop a rule library for basic optimizations.
\end{itemize}

The rest of the paper is structured as follows. Section~\ref{related} introduces the related work.
Section~\ref{pre} recalls the preliminary notations about quantum computing. 
Section~\ref{model} proposes a new representation of quantum circuits.
Section~\ref{qcr} discusses the design of the pattern matching-based quantum circuit rewriting framework.
Section~\ref{case} shows two case studies.
Section~\ref{exper} evaluates QRewriting by using the BIGD~\cite{Tan2020}, the Arithmetic and Toffoli~\cite{YSN2017} benchmarks.
Finally, Section~\ref{conclu} provides the conclusion. 

\section{Related Work}
\label{related} 
Several quantum circuit optimization compilers have recently been proposed to compile a quantum circuit to various processors.
For example, Qiskit \cite{qiskit} and t$|$ket$\rangle$ \cite{Sivarajah2020} support generic gate set;
Quilc \cite{markskilbeck} is tailored for the Rigetti Agave quantum processor.
There are several optimizers that automatically discover patterns~\cite{Zhihao2019, Kissinger2020, Pointing2021, xu2022quartz}.
QRewriting aims to rewrite quantum circuits
between different processors according to a given rule set, mainly focusing on pattern matching and replacement.

Pattern matching is widely used in circuit optimization.
For example, many algorithms have employed peephole optimization and pattern matching to optimize circuits.
Peephole optimization identifies small sets of instructions and replaces them with equivalent sets that have better performance~\cite{Liu21, McKeeman65}.
Exact matching is only feasible for small and medium-scale circuits~\cite{Abdessaied13}. 
Heuristics are often used in large-scale circuits, but they cannot ensure optimal results~\cite{Raban20, Rahman12}.  
In~\cite{Aditya06, Soeken16}, Prasad et al. and Soekens et al.
showed how to find  optimal quantum circuits for all 3-qubit functions.
Nam et al. proposed five optimization subroutines~\cite{YSN2017}.
Murali et al.  developed the first multi-vendor quantum computer compiler
which compiles from high-level languages to multiple real-system quantum computer prototypes,
with device-specific optimizations~\cite{Prakash2019full}.
The work of Chen et al.~\cite{chen2021} is the closest to ours,
where a quantum circuit optimization framework based on pattern matching (PaF) is proposed.
It uses subgraph isomorphism to find a pattern circuit 
in the target quantum circuit according to a given external rule description,
then replaces it with an equivalent one.

Previous work often treats a target circuit as a DAG, which is not able to identify the patterns that consist of several gates that appear non-consecutively.
In this paper, we introduce a new representation of  quantum circuits,
which can deal with non-consecutive patterns more conveniently.
For quantum circuit rewriting, we propose a polynomial-time algorithm,
which is based on dynamic programming to match and replace pattern circuits in the target circuit.

\section{Preliminary}
\label{pre}
In this section, we introduce some notions and notations of quantum computing.
Let $\mathbb{Z}$ and $\mathbb{C}$ denote the sets of all integers and complex numbers, respectively.

Classical information is stored in bits, while quantum information is stored in qubits. 
Besides two basic states $\ket{0}$ and $\ket{1}$, 
a qubit can be in any linear superposition state like $\ket{\phi}=a\ket{0}+b\ket{1}$, 
where $a, b\in \mathbb{C}$ satisfy the condition $|a|^{2}+|b|^{2}=1$.
The intuition is that $\ket{\phi}$ is in the state $\ket{0}$ with  probability $|a|^{2}$ and in the state $\ket{1}$ with probability $|b|^{2}$.
        
A quantum gate acts on a collection of qubits, which are called the operation qubits of the gate.
For example, the Hadamard gate (H gate) is applied on one qubit, and the CX gate is applied on two qubits.
Its behaviour is described as:
$$
{\rm CX}(\alpha\ket{0}\ket{\psi}+\beta\ket{1}\ket{\phi})=\alpha\ket{0}\ket{\psi}+\beta\ket{1}({\rm X} \ket{\phi}), 
$$ 
that is, we apply the X gate to the second qubit –- called the \textit{target} –- if the first –- the \textit{control} –- is in the state $\ket{1}$,
and the identity transformation otherwise, where $\ket{\psi}$ and $\ket{\phi}$ are the state of the second qubit. 
Two other gates which are relevant include the 3-qubit Toffoli gate CCX and the doubly-controlled phase gate CCZ.
Likewise, the CCX and CCZ gates apply X and Z gate, respectively, when the first two qubits are in state $\ket{1}$. 
Fig.~\ref{common_gates} lists the symbols of some commonly used quantum gates.

In a quantum circuit each line represents a \textit{wire}.
The wire does not necessarily correspond to a physical wire,
but may correspond to the passage of time or a physical particle that moves from one location to another through space.
The interested reader can find more details of these gates from many textbooks such as \cite{Nielsen2016}.
The execution order of a quantum logical circuit is from left to right.
The width of a quantum circuit refers to the number of qubits in the quantum circuit.
The depth of a quantum circuit refers to the number of layers executable in parallel.
We refer to a quantum circuit with depth less than 100 as a small-scale circuit,
a quantum circuit with depth more than 1000 as a large-scale circuit,
and the rest are medium-scale circuits.
\input{sections/commongate.tex}

\section{Circuit Representation}
\label{model}
In this section, we define a new representation of quantum circuits and the pattern matching condition,
which easily identifies the patterns that appear non-consecutively but are reducible.
Based on that, we will state the quantum circuit rewriting problem  considered in the paper.
\begin{definition}
\label{def_ssr}
An  instruction  is represented by a triple $(\gamma, \rho, \alpha)$, where 
\begin{itemize}
    \item $\gamma$ is the symbol of a gate type;
    \item $\rho$ is a finite sequence of operation qubits for a gate;
    \item $\alpha$ is a finite sequence of rotation angles for a gate;
\end{itemize}
\end{definition}
A quantum circuit $\Cc$ is a sequence of triples $(\gamma_0, \rho_0, \alpha_0)(\gamma_1, \rho_1, \alpha_1) \cdots$
$(\gamma_n, \rho_n, \alpha_n)$, and
the length of the sequence is denoted by $|\Cc|$.
The gate sequence $\Gamma_\Cc$ of  the quantum circuit $\Cc$ is
a symbol sequence of gate types obtained by projecting each element of $\Cc$ to its first component, e.g., $\Gamma_\Cc = ``\gamma_0\gamma_1\cdots\gamma_n$''. 
The new representation of a rule is a pair $\Pat=(\Cc_{p}, \Cc_s)$,
consisting of a pattern circuit $\Cc_{p}$ and a substitution circuit $\Cc_s$.
For simplicity, if the sequence is empty, we ignore it.
Table~\ref{gate_alias} lists the gate symbols and the distinct aliases.

\begin{table}
    \centering
        \caption{\label{gate_alias} The symbols of  gates and  the distinct aliases.}\vspace{1mm}
 \renewcommand\arraystretch{1.2}
     \label{tab:my_label}
\begin{tabular}{p{1cm}<{\centering}p{1.5cm}<{\centering}|p{1cm}<{\centering}p{1.5cm}<{\centering}|p{1cm}<{\centering}p{1.5cm}<{\centering}|p{1cm}<{\centering}p{1.5cm}<{\centering}}
					\hline
 Gates &Aliases & Gates& Aliases   & Gates& Aliases& Gates & Aliases  \\
\hline
I & `I' 
 & H & `h' & X  &`x' & Y & `y' \\ 
 Z  &`z'  &T& `t'  
& T$^{\dagger}$ &`T' & S &`s' \\
S$^{\dagger}$  & `S'&  {\rm R$_{\rm x}$}   
&`X' & {\rm R$_{\rm y}$}&`Y' 
& {\rm R$_{\rm z}$}&`Z' \\
 CX &`c'
& CZ  &`C'
& CCZ &`E'
 & CCX &`F'\\
\hline
\end{tabular}
\end{table}

\begin{example}
    \label{ex:1}
    \input{sections/ex1}
    Suppose we want to rewrite the quantum circuit $\Cc_t$ in Fig.~\ref{fig_ex1} (a).
    The new representation of the target quantum circuit $\Cc_t$  is
    \begin{itemize}
        \item $\Cc_t$ = (`x', [q[2]])(`x', [q[2]])(`c', [q[0], q[1]])(`c', [q[0], q[2]])(`c', [q[0], q[1]])(`x', [q[2]])(`x', [q[0]]),
    \end{itemize}
    and its gate sequence is represented by $\Gamma_t$ = ``xxcccxx''.
    We can make use of the rule set \{$\Pat_1$, $\Pat_2$, $\Pat_3$, $\Pat_4$\} in Fig.~\ref{fig_Xpattern} to help with the circuit rewriting, where
    \begin{itemize}
        \item $\Pat_1$ =  ((`x', [q[0]])(`x', [q[0]]), (`I', [q[0]])); 
        \item $\Pat_2$ = ((`c', [q[0], q[1]])((`c', [q[0], q[1]]), (`I', [q[0]])(`I', [q[1]])); 
        \item  $\Pat_3$ = ((`c', [q[0], q[1]])(`c', [q[1], q[2]])(`c', [q[0], q[1]]), (`c', [q[0], q[2]])\\
            (`c', [q[1], q[2]]));
        \item  $\Pat_4$ =  ((`x', [q[1]])(`c', [q[0], q[1]])(`x', [q[1]]), (`c', [q[0], q[1]]));
           \end{itemize}
           
   \input{sections/pattern_x_cx}
\end{example}

To facilitate the description of pattern matching, we introduce the following definition.

\begin{definition}
\label{def_subsequence}
Let $\Gamma$  and $\Gamma'$ be two sequences. We say  
 $\Gamma'$ is a subsequence of  $\Gamma$,
if there exist indices $0 \leq i_0 <\cdots< i_{|\Gamma'|-1} \leq |\Gamma|$ such that $\Gamma[i_{k}] = \Gamma'[k]$ for all $k \in [0, |\Gamma'|-1]$.
\end{definition}

The subsequence set is a set of distinct subsequences of the pattern circuit in the target circuit.
Note that we do not distinguish the indices from the gates to which the indices correspond in the  quantum circuit.
Suppose $\Gamma$ is a gate sequence.
Then $\Gamma[a:b]$ means to take the segment of $\Gamma$ from index $a$ to $b$ in,
if $b$ is not specified, we mean to take the suffix of $\Gamma$ from index $a$. 
Suppose that $s$ is a subsequence of $\Gamma$,
we write $\Gamma \setminus s$ to mean the subsequence of $\Gamma$ obtained by removing all occurrences of $s$.

  \begin{example}
    \label{ex:2}
   In Example~\ref{ex:1}, the gate sequences and its subsequence sets of the pattern circuits in the rule set are given as follows.
    \begin{itemize}
		\item ``xx'':  \{\{0, 1\}\};
		\item ``cc'':  \{\{2, 4\}\};
		\item ``ccc'':  \{\};
		\item ``xcx'':  \{\{1, 3, 5\}\}.
	\end{itemize}
\end{example}

\begin{definition}
\label{def_qmap}
(Qubit mapping) Given two qubit sets $Q$ and $Q'$, a qubit mapping function $f$ is a bijective function between the qubit sets $Q$ and $Q'$.
\end{definition}

\begin{definition}
\label{def_state}
(Qubit state independence) Let $\Cc_t$, $\Cc_p$ be two circuits with gate sequences  $ \Gamma_t$, $\Gamma_p$, respectively.
Suppose a subsequence $s=$\{$i_0$, $\ldots$, $i_l$\} of $\Gamma_t$ that can match  $\Gamma_p$.
We say the qubit state in $s$ is independent w.r.t. $\Gamma_t[i_0:i_l] \setminus s$,
if the control qubit set in $s$ 
does not intersect with the target qubit set of the gates in $\Gamma_t[i_0:i_l] \setminus s$, and vice-versa.
\end{definition}

\begin{definition}
\label{def_matche}
(Pattern matching)
Let $\Cc_t$ and $\Cc_p$ be a target circuit and a pattern circuit with gate sequences $ \Gamma_t$, $\Gamma_p$, respectively.
We say $\Cc_p$ matches $\Cc_t$ if the following two conditions hold:
\begin{itemize}
    \item  $\Gamma_t$ has a subsequence that can match $\Gamma_p$ up to a qubit mapping;
    \item  the qubit sets of the subsequence and the pattern circuit satisfy the qubit mapping and the qubit state independence conditions.
\end{itemize}
\end{definition}

\begin{example}
We continue the last example to show the difference between the new representation of quantum circuits and the DAG representation. 
Suppose the instructions of circuit $\Cc_t$ (resp. pattern circuit of $\Pat_2$) from left to right are named $g_0$--$g_6$ (resp. $g_0'$--$g_1'$).
In Fig.~\ref{dag}, (a) is the DAG representation of the  circuit segment $g_2$--$g_4$ and 
(b) is the DAG representation of the pattern circuit of $\Pat_2$. 
We can intuitively see that (a) has no subgraph isomorphic to (b).
But the pattern circuit of $\Pat_2$ matches the instructions $g_2$ and $g_4$. It satisfies the qubit mapping function \{$f$(q[0]) = q[0], $f$(q[1]) = q[1]\} and the qubit state independence condition.
Therefore,  the  circuit  $\Cc_t$  can be rewritten into  (`x', [q[2]])(`x', [q[2]])(`c', [q[0], q[2]])(`x', [q[2]])(`x', [q[0]]).
\end{example}
\input{sections/dag}

\begin{definition}
For a given  target circuit $\Cc_t$ and a rule $\Pat=(\Cc_p, \Cc_s)$,
a replacement candidate is a triple $(s, p, c)$, where 
	\begin{itemize}
		\item $s$ is a subsequence set of the target circuit $\Cc_t$ that can match the pattern circuit $\Cc_p$;
		\item $p$ is a rule $\Pat$;
		\item $c \in \mathbb{Z}$ is a conflict index, with the default value being $-1$.
	\end{itemize}
   \end{definition}

\begin{definition}
A target circuit has a replacement conflict if an index of the target circuit appears more than once in the subsequence set.
\end{definition}

The replacement candidates for a replacement conflict form a conflict set. A replacement scheduler is a set of replacement candidates for different indices.
\begin{example}
    Continuing the last example, we see that
    the subsequence set of the sequence  ``cc'' is \{\{2, 4\}\}, which appears non-consecutively in the sequence $\Gamma_t$.
    The first instruction (`x', [q[2]]) appears in both the subsequence sets of ``xx'' and ``xcx'', which means that in the target circuit different rules may be matched at the same index.
    Two schedulers are given as follows.
        \begin{itemize}
            \item s$_1$: \{(\{0, 1\}, $\Pat_1$, 1), (\{2, 4\}, $\Pat_2$)\},
            \item s$_2$: \{(\{1, 3, 5\}, $\Pat_4$, 1), (\{2, 4\}, $\Pat_2$)\}.
        \end{itemize}
    The scheduler $s_1$ (resp. $s_2$)  replaces the instructions in the index set \{0, 1\} (resp. \{1, 3, 5\}) using the substitution circuit of $\Pat_1$ (resp. $\Pat_4$). 
   After one of the schedulers is applied, we obtain the circuit in Figs.~\ref{fig_ex1} (b) or (c).
   Different schedulers result in different gate counts or depths of the rewritten circuits. The circuits $\Cc_t$ rewritten using schedulers s$_1$ or s$_2$ have the same gate count but their depths are  2 and 3 respectively.
   In both schedulers, the first element has the component $1$, which is an index to indicate where the conflict takes place.
   \end{example}

We are now ready to state the following problem.

\begin{problem}
\label{prob_migrate}
Given two gate sets $G_1$, $G_2$ and a rule set  that expresses the equivalence of $G_1$ by the elements of $G_2$,
how to rewrite a quantum circuit supported by gate set $G_1$  to a quantum circuit supported by $G_2$?
\end{problem}

By using our new representation of quantum circuits, we reduce the above problem
to finding distinct subsequences of the pattern sequence in the target sequence
up to a qubit mapping function and we use a 
qubit state independence condition to filter the obtained subsequences.


\section{Quantum Circuit Rewriting}
\label{qcr}
We propose a pattern matching-based quantum circuit rewriting framework. 
It consists of two steps. One matches the pattern circuit in the target circuit, the other replaces it. 

\subsection{Pattern Matching Algorithm}
\label{pm}
We propose an algorithm  based on dynamic programming to match
the patterns in a rule set against a target circuit. 
Let $\Cc_t$ and $\Cc_p$ be the target and pattern circuits with gate sequences
$\Gamma_t$ and $\Gamma_p$, respectively.
We consider  the problem of finding the distinct subsequences of the pattern sequence $\Gamma_p$ in the target sequence $\Gamma_t$. 
The obtained subsequences only match the gate types, so we also need to check whether the operation qubits in the subsequences  satisfy the qubit mapping function and qubit state independence condition.

The input of Algorithm~\ref{algo_position} is a target circuit $\Cc_t$ and a rule set $\Pat$, and the output is a set of replacement candidates $\M$.
The function $distinct\_subsequence(\Gamma_t, \Gamma_p, \delta)$ uses  a dynamic programming algorithm to compute the distinct subsequences of $\Gamma_t$ that can match $\Gamma_p$ and returns the subsequence set to $\D$.
The function $check\_qubit\_condition(\D, \Cc_t, p)$ checks whether the results in $\D$ satisfy the qubit mapping function and qubit state independence condition. 

The input of the function $distinct\_subsequence(\Gamma_t, \Gamma_p, \delta)$ is a target sequence $\Gamma_t$, a pattern sequence $\Gamma_p$, and a parameter $\delta$ to limit the range of indices of $\Gamma_p$ in $\Gamma_t$. The symbol $s[0]$ denotes the first component of the replacement candidate $s$.
The output is a set $D[m+1]$ recording the distinct subsequences of $\Gamma_t$ that can match $\Gamma_p$. 
We use the set
$D[j+1]$ to record the subsequences of  $\Gamma_t$ that can match $\Gamma_p[0:j]$.
If the condition  $\Gamma_t[i]=\Gamma_p[j]$ is satisfied, there are subsequences of $\Gamma_t[0:i]$ that can match $\Gamma_p[0:j]$. 
Line 8 updates the set $D[j+1]$. 
To find the subsequence of $\Gamma_p[0:j]$, we need to 
first calculate the subsequences of $\Gamma_p[0:j-1]$. 
The update $D[j+1] \leftarrow D[j+1]\cup \{s\cup\{i\}: s \in D[j] \ {\rm and }\  j-s[0]<\delta\}$ is the Bellman equation~\cite{DP2003}, which is a necessary enabler of the dynamic programming algorithm. 

The time complexity of Algorithm~\ref{algo_position} is $\mathcal{O}(kmn)$, where $k$ is the number of rules in the rule set $\Pat$, $m$ is the maximum length of the gate sequence $\Gamma_p$ in the rule set $\Pat$, and $n$ is the length of the gate sequence $\Gamma_t$. The space complexity is $\mathcal{O}(m)$.

\input{sections/algo1}

\begin{example}
Let us consider the  quantum circuit in Fig.~\ref{fig_16qbt} (a) and the rule set $\Pat=\{\Pat_1, \Pat_2, \Pat_3, \Pat_4\}$ in Fig.~\ref{fig_Xpattern}.
The gate sequences of the target circuit and the pattern circuits are ``xxxxcccxxxxxcccxxxxcccxxxcccxxxxxxccc'', ``xx'', ``cc'', ``ccc'', and ``xcx'', respectively. 
The  set of subsequences of ``cc'' is  as  follows.
\begin{itemize}
    \item \{\{4, 5\}, \{4, 6\}, \{5, 6\}, $\cdots$\}.
\end{itemize}
The corresponding instructions of the indices \{4, 5, 6\}  in the target circuit $\Cc_t$ are (`c', [q[13], q[2]])(`c', [q[9], q[14]])(`c', [q[4], q[12]]).
The subsequences \{4, 5\}, \{4, 6\}, \{5, 6\} do not satisfy the qubit mapping function condition.
We use the function $check\_qubit\_condition()$ to filter the subsequences and finally get the subsequence set of ``cc'' in the gate sequence of the target circuit, which  is \{\{5, 12\}, \{19, 25\}\} highlighted with dotted lines in Fig.~\ref{fig_16qbt} (a).

\end{example}

\subsection{Replacement Algorithm}
\label{ss}
By Algorithm~\ref{algo_position}, we obtain all the subsequences of the pattern circuits in the target circuit. To resolve replacement conflicts, we propose three conflict resolution policies. They give rise to three variants of QRewriting called QPRewriting, QGRewriting, and QSRewriting, respectively.
Due to the decoherence of qubits, the lifetime of qubits is very short~\cite{2019Zhang}.
The execution time of the  quantum circuit is determined by several factors such as  the depth  and the gate count of the  quantum circuit.
Here, we mainly use the depth to select the optimal replacement scheduler.

\begin{itemize}
    \item Precise policy calculates all the candidates when a replacement conflict occurs.
    \item Greedy policy follows the principle ``first come, first served''. That is, it chooses the one that appears first in the target circuit among the conflict set.   
    \item Stochastic policy selects a candidate stochastically in the conflict set for the scheduler.
\end{itemize}
\input{sections/algo3.tex}

We propose an algorithm based on the breadth-first search to compute  the replacement scheduler as shown in Algorithm~\ref{algo_exact}.
The input is a gate  sequence $\Gamma_t$ and a set of replacement candidates $\M$.
$\Ss$ is a scheduler set, and the queue $Q$ stores the sub-scheduler.
Firstly, we push one element $\{(\{\}, \epsilon, -1)\}$ into $Q$. Then, we loop the queue $Q$ until it is empty in lines 4--13.
The function $next\_conflict(\Gamma_t, \M, s)$ computes the next conflict index $i$ in $\Gamma_t$ from the current conflict index to the end of $\Gamma_t$.
If there is no conflict at this index, we directly add it to  $s$, otherwise return the index.
 When arriving at the end of $\Gamma_t$, we add the scheduler $s$ into $\Ss$.
Line 10, according to the conflict policy,  calculates the candidate set that  has a conflict at index $i$ in $\M$.
Lines 11--13 append the replacement candidates to $s$ and push it into queue $Q$.
Finally, we calculate the depth of the replaced circuit and return the scheduler with the smallest depth.

The time complexity depends on the conflict policy. 
In the worst case, the precise policy is used and the time complexity is $\mathcal{O}(n^m)$, where $m$ is the number of conflicts and $n$ is the size of the conflict set.  When dealing with large-scale circuits, the precise policy is not scalable. Therefore we do not show the precise policy in our experiments.
The time complexity of both greedy and stochastic policies is $\mathcal{O}(mn)$.

\begin{example}
Let us continue the last example, where we show an example of generating schedulers. Starting from the index $i=0$, we search for the next conflicting index $i$ in $\M$.
When $i=18$, the  conflict set is \{(\{10, 18\}, $\Pat_1$, 18), (\{18, 26, 31\}, $\Pat_4$, 18)\}. If the precise policy is used, we append the scheduler $s$ with the two candidates and put into the queue $Q$, respectively, and the generated schedulers are given as follows. 

\begin{tabularx}{\linewidth}{lX}
$s_1$: & $s_2$:\\
\quad (\{0, 22\}, $\Pat_1$) & \quad (\{0, 22\}, $\Pat_1$)\\
\quad (\{2, 13, 17\}, $\Pat_4$)\quad \quad \quad &\quad (\{2, 13, 17\}, $\Pat_4$)  \\
\quad (\{3, 9\}, $\Pat_1$) & \quad (\{3, 9\}, $\Pat_1$) \\
\quad (\{5, 12\}, $\Pat_2$) &\quad (\{5, 12\}, $\Pat_2$)  \\
\quad (\{10, 18\}, $\Pat_1$, 18) &\quad  (\{18, 26, 31\}, $\Pat_4$, 18) \\
\quad (\{15, 23\}, $\Pat_1$) &\quad (\{15, 23\}, $\Pat_1$)  \\
\quad (\{16, 24\}, $\Pat_1$) &\quad (\{16, 24\}, $\Pat_1$)   \\
\quad (\{19, 25\}, $\Pat_2$) &\quad (\{19, 25], $\Pat_2$)  \\
\end{tabularx}
 If we use the greedy policy, the replacement candidate (\{10, 18], $\Pat_1$, 18) will be selected, and the generated scheduler is $s_1$. If we use the stochastic policy, one of them is selected  and the finally generated scheduler is either $s_1$ or $s_2$.
\end{example}

Algorithm \ref{algo_substitute} inputs a target circuit $\Cc_t$ and a scheduler $\Ss$, and outputs a replaced circuit. We reversely traverse each element of the scheduler $\Ss$. 
Line 2 obtains the mapping relationship $qmaps$ of qubits between the subsequence of the target circuit and the pattern circuit, according to the qubit mapping function $qubits\_mapping(\Ss[i])$. 
Lines 6--11 update the instructions on the target circuit with the substitution circuit one by one.
If the substitution gate sequence is longer than the pattern gate sequence,
the redundant gates are inserted after the index where the pattern circuit appears in the target circuit.
Lines 10--11 remove redundant locations for the target circuit.
The time complexity is $\mathcal{O}(mn)$, where $n$ is the length of the scheduler, and $m$ is the maximum length of the pattern circuit.

\input{sections/algo4}

\begin{example}
Continuing the last example, we consider the replacement candidate $\Ss[i]$= (\{2, 13, 17\}, $\Pat_4$) as an example.  We can have that 
\begin{itemize}
    \item $\Pat_4$ = ((`x', [q[0]])(`c', [q[0], q[1]])(`x', [q[1]]), (`c', [q[0], q[1]]));
    \item $qmaps=$ \{$f$(q[0]) = q[3], $f$(q[1]) = q[8]\}.
\end{itemize}
The length of the pattern circuit is greater than that of the substitution circuit.
First, we get $n=1$ and update the second instruction of the target circuit $\Cc_t$ with the instruction (`c', [q[3], q[8]]). Then we remove the 13th and 17th instructions of the target circuit.
\end{example}

\subsection{Quantum Circuit Optimization}
\label{qco}
We develop a rule library for basic optimizations.
To facilitate the distinction between internally optimized rules and circuit rewriting rules, 
we divide the library into an internal library and an external one. The external library is a rule set provided by a user, which can be a rule library for optimization or rewriting.
The internal  library is mainly used for basic reduction and quantum gate exchange~\cite{YSN2017,Iwama2002}. Note that almost all the gates implemented by quantum hardware devices are usually 1-qubit and 2-qubit gates, so our rule library mainly concerns  1-qubit gates and 2-qubit gates. 
The maximum input scale involved in the rule set is 3-qubit. The gate specification involves some cancellation rules for 1-qubit gates and  2-qubit gates, as shown in Fig.~\ref{redrule}.
The commutation rules shown in Fig.~\ref{comrule} include the  transformation rules given in \cite{Iwama2002}.

It is possible that  after a step of circuit rewriting the target circuit still matches some rules. We repeat several rounds of internal optimization and circuit rewriting until no  pattern circuit can be matched or the specified repetition bound is reached (5 by default in practice).

\input{sections/cancellation.tex}
\input{sections/commutation.tex}

\section{Case Study}
\label{case}
In this section we consider two examples: one rewrites a circuit with three CCZ gates to a circuit using the $G_{Sur}$ gate set; the other optimizes a circuit with a stochastic policy.

\subsection{Rewriting Circuits for Surface-17}
\label{case_tof_3}
\begin{figure}
    \centering
	\scalefont{1}
   	\begin{tikzpicture}
		\node at (5.5, 5){ 
		\Qcircuit @C=0.8em @R=0.4em @!R{
 		 \lstick{\rm q[0]}& \qw &\ctrl{4}& \qw&\qw &\qw&\ctrl{4}&\qw&\qw\\
 		\lstick{\rm q[1]} & \qw &\ctrl{3}& \qw&\qw&\qw &\ctrl{3}&\qw&\qw\\
 		 \lstick{\rm q[2]}& \qw &\qw & \qw &\ctrl{1} &\qw&\qw&\qw&\qw\\
 		 \lstick{\rm q[3]}&\qw &\qw &\gate{\rm H} &\gate{\rm Z} &\gate{\rm H}&\qw&\qw&\qw\\
 		 \lstick{\rm q[4]}&\gate{\rm H} &\gate{\rm Z} &\gate{\rm H} &\ctrl{-1} &\gate{\rm H} &\gate{\rm Z} &\gate{\rm H} &\qw\\
 		 }};
	\end{tikzpicture}
    \caption{The  quantum circuit Toff-NC$_3$.}
    \label{fig_sur_before}
\end{figure}
We demonstrate the use of QRewriting to rewrite the  quantum circuit TOF\_3~\cite{Tan2020} that has three occurrences of the CCZ gates to the $G_{Sur}$ gate set~\cite{LaoSAA22}.
The target circuit $\Cc_t=$(`h', [q[4]])(`E', [q[0], q[1], q[4]])$\cdots$ is displayed  in Fig. \ref{fig_sur_before}, 
and the rules are listed in Fig.~\ref{surf}.
The CCZ gate decomposition rule~\cite{Amy2018}  $\Pat=(\Cc_p, \Cc_s)$ is a pair, where 
\begin{itemize}
    \item  $\Cc_p=$ (`E', [q[0], q[1], q[2]]);
    \item   $\Cc_s=$ (`t', [q[0]])(`t', [q[1]]) $\cdots$.
\end{itemize}
The gate sequence of the target circuit and the decomposition pattern circuit are ``hEhhEhhEh'' and ``E'', respectively. 
The subsequence set of  $\Cc_t$ that can match ``E'' is \{\{1\}, \{4\}, \{7\}\}.
Finally, the resulting circuit is shown in Fig.~\ref{fig_sur_after}.
\input{sections/pattern_surf}
\input{sections/tof3_after}

\subsection{Optimization with a Stochastic Policy}
\label{case_opt}
Next we show an example of circuit optimization using QSRewriting with a stochastic policy.
The target circuit 
and a set of rules  $\Pat=\{\Pat_1, \Pat_2, \Pat_3, \Pat_4\}$
are  shown in Fig.~\ref{fig_16qbt} (a) and Fig.~\ref{fig_Xpattern}, respectively. 
The gate sequences of the target circuit and the pattern circuits are ``xxxxcccxxxxxcccxxxxcccxxxcccxxxxxxccc'', ``xx'', ``cc'', ``ccc'' and ``xcx'', respectively. 
The subsequence sets of the gate sequence of pattern circuits are given as follows, 
\begin{itemize}
    \item  ``xx'': 
\{\{3, 9\}, 
\{10, 18\}, 
\{0, 22\}, 
\{15, 23\}, 
\{16, 24\}\};
    \item  ``cc'': \{\{5, 12\}, \{19, 25\}\};
    \item  ``ccc'': \{\};
    \item  ``xcx'':  \{\{2, 13, 17\}, \{18, 26, 31\}\}.
\end{itemize}
The replacement candidates (\{10, 18\}, $\Pat_1$, 18) and (\{18, 26, 31\}, $\Pat_4$, 18)  have a conflict at the index $18$ of the target circuit.
With the stochastic policy, either of the candidates can be chosen. Suppose the former is taken, then the generated replacement scheduler is given as follows.
\begin{tabularx}{\linewidth}{lX}
$s_1$:\\
\quad (\{0, 22\}, $\Pat_1$)\\
\quad (\{2, 13, 17\}, $\Pat_4$) \\
\quad (\{3, 9\}, $\Pat_1$) \\
\quad (\{5, 12\}, $\Pat_2$)  \\
\quad (\{10, 18\}, $\Pat_1$, 18)  \\
\quad (\{15, 23\}, $\Pat_1$)  \\
\quad (\{16, 24\}, $\Pat_1$)   \\
\quad (\{19, 25\}, $\Pat_2$) \\
\end{tabularx}
Finally, we obtain the resulting circuit in Fig.~\ref{fig_16qbt} (b),  which reduces the gate count and the depth  by 48.65\% and  20\%, respectively.
\input{sections/16qbt}

\section{Experiments}
\label{exper}

We compare QRewriting  with the state-of-the-art algorithm for quantum circuit optimization framework based on pattern matching, namely PaF~\cite{chen2021}. Notice that PaF is not freely available, so we implemented that algorithm in Python.
The  implementation of QRewriting in Python is available at \url{https://github.com/ShepherdLee519/qcpm.git}.
All the experiments are conducted on a Ubuntu machine with a 2.2GHz CPU and 64G memory. 
For the stochastic policy, we execute QSRewriting five times and take the best result;  for other policies the executions are deterministic, so we execute them only once.

To compare with PaF, we adopt the BIGD benchmarks~\cite{Tan2020}, and use the rule set  shown in Fig.~\ref{fig_Xpattern}. 
We  use the gate count and the depth as our evaluation  metrics. 
The selected benchmarks are characterized by the parameters $(d_1, d_2)$, which is called gate density vector~\cite{Tan2020}. 
The two components stand for the densities of 1-qubit and 2-qubit gates in a benchmark. Suppose a quantum circuit has $n$ qubits, $M_{1}$ (resp. $M_{2}$) is the number of 1-qubit (resp. 2-qubit) gates, and the longest dependency chain is $l$, then $d_1 = M_1/(n \times l)$ and $d_2 = 2\times M_{2}/(n \times l)$. 

The BIGD benchmarks include 360 circuits with a total number of 129600 gates. 
After a PaF optimization, the gate count and the depth decrease by 66512 and 4009 within 6760 seconds. 
QSRewriting (resp. QGRewriting) takes 2816 (resp. 1982) seconds to rewrite these circuits, 
and the resulting circuits further reduce the 1-qubit gate,  2-qubit gates, total gate count and depth by 18.7\% (resp. 16.5\%), 13.9\% (resp. 12.8\%), 14.6\% (resp. 13.2\%), and 29.0\%   (resp. 26.2\%) compared with PaF.
Therefore, QRewriting is about three times faster than PaF, and the resulting circuits are better optimized.
The main evaluation results are shown in Fig.~\ref{spin_qrewriting} (a)--(d) which compare the performance of QSRewriting, QGRewriting and PaF in terms of 1-qubit gates, 2-qubit gates, the total gate counts and depths of the optimized circuits.
The cyan bar represents the gate count (depth) of the benchmarks.
The blue, red, and yellow colors are for PaF, QGRewriting, and QSRewriting, respectively.
We can see that the red and yellow lines are  mostly lower than the blue, the yellow is mostly obscured by the red, but we can still see that it is lower than the red in some places.
In Fig.~\ref{spin_qrewriting} (d), we can see that in a few cases the depth of the  quantum circuit might increase after some circuits are optimized.
The reason is that the gates of a rear layer may be moved to a front layer, causing the original gate of the front layer to conflict with it.

\input{sections/exp1}

Now we 
rewrite the Arithmetic and Toffoli benchmarks~\cite{YSN2017}, which contain 33 circuits and 201554 gates, from the $G_{IBM}$ gate set to the $G_{Sur}$ gate set (cf. Table~\ref{gateset}).
The Surface-17 processor limits 1-qubit gates to X and Y rotations, and more specifically $\pm\frac{\pi}{4}$, $\pm\frac{\pi}{2}$, and $\pm\pi$ degrees will be used in our decomposition. The primitive 2-qubit gate on this processor is CZ~\cite{LaoSAA22}.
In this experiment, we simply choose the greedy policy since no replacement conflict arises.

In Tables~\ref{tof_table_num} and~\ref{tof_table_dep}, we list the experimental data. Comparing the circuit rewriting with and without internal optimization, the gate count is reduced by up to 52\%, and the depth is reduced by up to 49\%. However, there is a price to pay. For a  quantum circuit with millions of instructions, a rewriting without optimization takes about 15 minutes, while a rewriting with optimization may take about two hours, depending on the size of the internal rule library.
\input{sections/tabs}

\section{Conclusion}
\label{conclu}

We introduced a new representation of quantum circuits,
which reduced the pattern matching of circuits to the problem of finding distinct subsequences.
We presented an algorithm based on dynamic programming to match the pattern circuits in the target circuit.
To resolve replacement conflicts,
we proposed three policies for generating replacement schedulers and a polynomial-time replacement algorithm.
We developed a rule library for basic optimizations
and applied it to rewrite the Arithmetic and Toffoli benchmarks from the $G_{IBM}$ gate set to the $G_{Sur}$ gate set.
Compared with the existing tool PaF, QRewriting improved the depth (resp. gate count) reduction by 29\% (resp. 14\%),
which demonstrated the effectiveness of our approach.

%
%
\bibliographystyle{splncs04}
\bibliography{mybibliography}
\end{sloppypar}
\end{document}

%% file: sections/commongate.tex
\begin{figure}[htbp]
	 \setlength{\arraycolsep}{3pt}
	 \begin{center}
	\begin{tikzpicture}
\node at (0, 8){I gate };
\node at (3, 8){	
	\Qcircuit @C=2.2em @R=1.75em {
	 & \gate{\rm I}  & \qw   \\						 
}};
\node at (6, 8){X gate };
\node at (9, 8){	
	\Qcircuit @C=2.2em @R=1.75em {
	 & \gate{\rm X}  & \qw    \\						 
}};
	\node at (0, 7){	H gate };
	\node at (3, 7){	
	\Qcircuit @C=2.2em @R=1.75em {
		 & \gate{\rm H}  & \qw 	 \\  						 
	}};
\node at (6, 7){	Y gate };
\node at (9, 7){	
	\Qcircuit @C=2.2em @R=1.75em {
	 & \gate{\rm Y}  & \qw 	 \\						 
}};

\node at (0, 6){	Z gate };
\node at (3, 6){	
	\Qcircuit @C=2.2em @R=1.75em {
	 & \gate{\rm Z}  & \qw 	 \\						 
}};

\node at (6, 6){S gate };
\node at (9, 6){	
	\Qcircuit @C=2.2em @R=1.75em {
	 & \gate{\rm S}  & \qw 	 \\						 
}};
\node at (0, 5){	T gate };
\node at (3, 5){	\Qcircuit @C=2.2em @R=1.75em {
	 & \gate{\rm T}  & \qw 	 \\						 
}};
	
\node at (6, 5){\rm R$_{\rm z}(\theta)$ gate };
\node at (9, 5){	\Qcircuit @C=2.2em @R=1.75em {
	 & \gate{\rm R_{z}(\theta)} & \qw 	 \\						 
}};
		 
	\node at (0, 4){	CX gate };
	\node at (3, 4){	
		\Qcircuit @C=2.2em @R=1.25em {
		 & \ctrl{1}  & \qw 	 \\
		 &\targ  	 & \qw   \\	    			}};
	\node at (6, 4){	CZ gate };
	\node at (9, 4){	
		\Qcircuit @C=2.2em @R=1.25em {
		 & \ctrl{1}  & \qw 	 \\
		 &\ctrl{-1} 	 & \qw   \\	    			}};
	\node at (0, 2.5){	CCX gate };
	\node at (3, 2.5){	
		\Qcircuit @C=2.2em @R=1.25em {
		 & \ctrl{2}  & \qw 	 \\
		 & \ctrl{1}  	 & \qw   \\	   
		 &\targ  	 & \qw   \\	   			}};
	
	\node at (6, 2.5){	CCZ gate };
	\node at (9, 2.5){	
		\Qcircuit @C=2.2em @R=1.25em {
		 & \ctrl{2}  & \qw 	 \\
		 & \ctrl{1}  	 & \qw   \\	   
		 &\gate{\rm Z}  	 & \qw   \\	   			}};
\end{tikzpicture}
\end{center}
\caption{Symbols of some commonly used  quantum gates.}
\label{common_gates}
\end{figure}

%% file: sections/ex1.tex
\begin{figure*}
  \centering
  \subfigure[]{
  		\begin{tikzpicture}
		\node at (5.5, 5){ 
			\Qcircuit @C=0.8em @R=0.4em @!R{
             \lstick{\rm q[0]}&\ctrl{1}&\qw &\ctrl{2} &\ctrl{1}&\gate{\rm X}&\qw\\
            \lstick{\rm q[1]} &\targ &\qw &\qw &\targ&\qw&\qw\\
             \lstick{\rm q[2]}&\gate{\rm X}&\gate{\rm X}&\targ &\gate{\rm X}&\qw&\qw\\
 }
		};
	\end{tikzpicture}\quad\quad
  }
  \subfigure[]{
  		\begin{tikzpicture}
		\node at (5.5, 5){ 
			\Qcircuit @C=0.8em @R=0.4em @!R{
             \lstick{\rm q[0]}&\ctrl{2}&\gate{\rm X}&\qw\\
             \lstick{\rm q[1]}&\qw&\qw&\qw\\
            \lstick{\rm q[2]} &\targ&\gate{\rm X}&\qw\\
 }
		};
	\end{tikzpicture}\quad\quad
  }
  \subfigure[]{
  		\begin{tikzpicture}
		\node at (5.5, 5){ 
			\Qcircuit @C=0.8em @R=0.4em @!R{
           \lstick{\rm q[0]}&\qw&\ctrl{2}&\gate{\rm X}&\qw\\
            \lstick{\rm q[1]} &\qw&\qw&\qw&\qw\\
          \lstick{\rm q[2]} &\gate{\rm X}  &\targ&\qw&\qw\\
 }
		};
	\end{tikzpicture}
  }
  \caption{(a) The  quantum circuit. (b) and (c) are the results of rewriting quantum circuit (a) using schedulers $s_1$ and $s_2$, respectively.} 
  \label{fig_ex1}
\end{figure*}

%% file: sections/pattern_x_cx.tex
\begin{figure}[htb]
	\begin{center}
	\scalefont{1}
	\begin{tikzpicture}
		\node at (5.5, 0){ 
			\Qcircuit @C=0.5em @R=0.4em {
			&&&&&&&&&&  &&& 
			&\ctrl{2}&\qw&\qw&\ctrl{2}&\qw&&& \gate{\rm I}&\qw&& && 
			\\
            &\gate{\rm X}&\gate{\rm X}&\qw & &\push{\rule{.1em}{0em}=\rule{.1em}{0em}}&&& \gate{\rm I}&\qw  &&&& 
            &&&&& &\push{\rule{.1em}{0em}=\rule{.1em}{0em}}&&&&&&&
            \\
            &&&&&&&&&&  &&& 
            &\targ&\qw&\qw&\targ&\qw&&& \gate{\rm I}&\qw&& &&
	}
		};
	\end{tikzpicture}
		\begin{tikzpicture}
		\node at (5.5, 0){ 
			\Qcircuit @C=0.5em @R=0.4em{
			&\ctrl{1}&\qw&\ctrl{1}&\qw && &\ctrl{2}&\qw&\qw&\qw &&&& 
			&\qw&\ctrl{2}&\qw&\qw &&& \ctrl{2}&\qw
			\\
            &\targ& \ctrl{1}&\targ&\qw &\push{\rule{.1em}{0em}=\rule{.1em}{0em}}&&  \qw&\qw&\ctrl{1}&\qw  &&&& 
            &&&& &\push{\rule{.1em}{0em}=\rule{.1em}{0em}}& 
            \\
            &\qw&\targ&\qw&\qw &&   &\targ&\qw&\targ&\qw  &&&& 
           & \gate{\rm X} &\targ & \gate{\rm X}&\qw  &&   &\targ &\qw
	}
		};
	\end{tikzpicture}
	\end{center}
	\caption{The rules used to optimize the X and
CX gates.}
	\label{fig_Xpattern}
\end{figure}

%% file: sections/dag.tex
\begin{figure}
    \centering
        \begin{tikzpicture}
         \node at (1.5, -1){(a)};
         \node at (5, -1){(b)};
        \node at (1.25, 1.25){$g_2$};
        \draw [black, thin] (1.25, 1.25) circle [radius=0.4];
        \node at (2.5, 0){$g_3$};
        \draw [black, thin] (2.5, 0) circle [radius=0.4];
        \node at (2.5, 2.5){$g_4$};
        \draw [black, thin] (2.5, 2.5) circle [radius=0.4];
       
       \draw [->, thin] (1.55, 0.9) -- (2.2, 0.35);
       \draw [->, thin] (1.55, 1.6) -- (2.15, 2.2);
       \draw [->, thin] (2.5, 0.45) -- (2.5, 2.05);
       
        \node at (1.65, 0.45){q[0]};
        \node at (1.55, 2.05){q[1]};
        \node at (2.85, 1.25){q[0]};
        
        \node at (4, 1.25){$g_0'$};
        \draw [black, thin] (4, 1.25) circle [radius=0.4];
        \node at (6, 1.25){$g_1'$};
        \draw [black, thin] (6, 1.25) circle [radius=0.4];
        \draw [->, thin] (4.45, 1.35)  -- (5.55, 1.35);
        \draw [->, thin] (4.45, 1.15)  -- (5.55, 1.15);
        \node at (4.9, 1.6){q[0]};
        \node at (4.9, 0.9){q[1]};
        \end{tikzpicture}
        \caption{
        (a) and (b) are the DAG representations of the circuit segment $g_2$--$g_4$ of $\Cc$ and the pattern circuit of $\Pat_2$, respectively.}
        \label{dag}
    \end{figure}

%% file: sections/algo1.tex
\begin{algorithm}[htb]
	\caption{$pattern\_matching(\Cc_t$, $\Pat$)}	
	\label{algo_position}
	\LinesNumbered  
	\KwIn{a quantum circuit $\Cc_t$ and a rule set $\Pat$;}
	\KwOut{a set of substitution candidate $\M$;}
	$\M \gets \emptyset$; \\
	$\Gamma_t \gets$ the gate sequence of circuit $\Cc_t$;\\
	\ForEach{$p \in \Pat$}{
			$\Gamma_p \gets$ the gate sequence of pattern circuit of $p$;\\
			$\D \gets distinct\_subsequence(\Gamma_t, \;\Gamma_p, \; \delta)$;\\
			\If{$check\_qubit\_condition(\D, \; \Cc_t, \; p)$}{
			$\M \gets \M \cup\{(\D, \; p, \; -1)\};$
			}
	}
	\Return $\M$;
	\end{algorithm}
	
\begin{algorithm}[htb]
		\caption{$distinct\_subsequence(\Gamma_t, \;\Gamma_p, \; \delta) $ (Bellman \cite{DP2003})}
		\label{algo_dynamic}
		\LinesNumbered  
		\KwIn{two sequences $\Gamma_t$ and $\Gamma_p$, and a parameter $\delta$;}
		\KwOut{a set of subsequences of  $\Gamma_p$;}
		$n \gets $ the length of $\Gamma_t$;\\
		$m \gets $ the length of $\Gamma_p$;\\
		let $D$ be an array of length $m+1$;\\
		$D[0]\gets \emptyset$;\\
		\For{$i \gets 0$ to $n$}{
			\For{$j \gets \min(i,m)$ to $0$}{
				\If{$\Gamma_t[i]=\Gamma_p[j]$}{
				    $D[j+1]\gets D[j+1] \cup \{s\cup \{i\}: s\in D[j]\  {\rm and }\  j-s[0]<\delta \}$; \\
						}
					}
				}
		\Return $D[m+1]$;
		\end{algorithm}

%% file: sections/algo3.tex
\begin{algorithm}[htb]
	\caption{solve\_conflicts($\Gamma_t, \M$)}  
	\label{algo_exact}
	\LinesNumbered  
	\KwIn{a gate sequence $\Gamma_t$ and a set  of  replacement candidates  $\M$;}
	\KwOut{a replacement scheduler;}
	$\Ss \gets \emptyset$; \\
    let $Q$ be a scheduler queue; \\
    $Q.push(\{(\{\}, \epsilon, -1)\})$; \\
	\While{$Q$ is not empty}{
	    $s\gets Q.pop()$; \\
	    $i \gets next\_conflict(\Gamma_t, \M, s)$;\\
	    \If{i is None}{
	    $\Ss \gets \Ss \cup \{s\}$;\\
	    \textbf{continue};
	    }
	    $\mathcal{N} \gets$ Compute the conflict set on index $i$;\\
	    \ForEach{$c \in \mathcal{N}$}{
		$s\gets s\cup \{c\}$;\\
	    $Q.push(s)$;\\
	    }
	}
	\Return $compute\_depth(\Ss$); 
	\end{algorithm}

%% file: sections/algo4.tex
\begin{algorithm}[htb]
	\caption{substitute($\Cc_t, \Ss$)}  
	\label{algo_substitute}
	\LinesNumbered  
	\KwIn{
	a quantum circuit $\Cc_t$, and a substitution scheduler $\Ss$;\\
	}
	\KwOut{the substituted circuit $\Cc_t$;}
	\For{i $\gets$ length($\Ss$)-1 to 0}{
	$qmaps \gets qubits\_mapping(\Ss[i])$;\\
	$l_1 \gets$ the length of the pattern circuit of $\Ss[i]$; \\
	$l_2 \gets$ the length of the substitution circuit of $\Ss[i]$;\\
	$n \gets min(l_1, l_2)$;\\
	\For{j $\gets$ 0 to n-1}{
	$\Cc_t.update(\Ss[i], j, qmaps)$;
	}
	\If{$l_2$ > n}{
	$\Cc_t.insert\_gate(\Ss[i], qmaps, n);$
	}
	\If{$l_1$ > n}{
	$\Cc_t.delete\_gate(\Ss[i], n);$\\
	}
	}
	return $\Cc_t;$
	\end{algorithm}

%% file: sections/cancellation.tex
\begin{figure}[htb]
	\begin{center}
	\scalefont{1}
		\begin{tikzpicture}
		\node at (5.5, 5){ 
			\Qcircuit @C=0.5em @R=0.4em {
            &\gate{\rm R_x(-\theta_1)}& \gate{\rm R_x(\theta_1)}&\qw &\push{\rule{.1em}{0em}=\rule{.1em}{0em}}&&\gate{\rm I} &\qw 
          && 
            &\gate{\rm R_x(\theta_1)}& \gate{\rm R_x(\theta_2)}&\qw &\push{\rule{.1em}{0em}=\rule{.1em}{0em}}&&\gate{\rm R_x(\theta_1+\theta_2)} &\qw 
 }
		};
	\end{tikzpicture}
		\begin{tikzpicture}
		\node at (5.5, 5){ 
			\Qcircuit @C=0.5em @R=0.4em {
            &\gate{\rm R_y(-\theta_1)}& \gate{\rm R_y(\theta_1)}&\qw &\push{\rule{.1em}{0em}=\rule{.1em}{0em}}&&\gate{\rm I} &\qw 
          && 
            &\gate{\rm R_y(\theta_1)}& \gate{\rm R_x(\theta_2)}&\qw &\push{\rule{.1em}{0em}=\rule{.1em}{0em}}&&\gate{\rm R_x(\theta_1+\theta_2)} &\qw 
 }
		};
	\end{tikzpicture}
	\begin{tikzpicture}
		\node at (5.5, 5){ 
			\Qcircuit @C=0.5em @R=0.4em {
            &\gate{\rm R_z(-\theta_1)}& \gate{\rm R_z(\theta_1)}&\qw &\push{\rule{.1em}{0em}=\rule{.1em}{0em}}&&\gate{\rm I} &\qw 
          && 
            &\gate{\rm R_z(\theta_1)}& \gate{\rm R_z(\theta_2)}&\qw &\push{\rule{.1em}{0em}=\rule{.1em}{0em}}&&\gate{\rm R_z(\theta_1+\theta_2)} &\qw 
 }
		};
	\end{tikzpicture}
	\begin{tikzpicture}
		\node at (5.5, 5){ 
			\Qcircuit @C=0.5em @R=0.4em {
            &\gate{\rm S}& \gate{\rm S}&\qw &\push{\rule{.1em}{0em}=\rule{.1em}{0em}}&&\gate{\rm Z} &\qw 
          && 
          \gate{\rm H}&\gate{\rm X}&\gate{\rm H} &\push{\rule{.1em}{0em}=\rule{.1em}{0em}}&& \gate{\rm Z}&\qw 
          &&
           \gate{\rm H}&\gate{\rm Y}&\gate{\rm H} &\push{\rule{.1em}{0em}=\rule{.1em}{0em}}&& \gate{\rm Y}&\qw 
            &
	}
		};
	\end{tikzpicture}
	\begin{tikzpicture}
		\node at (5.5, 5){ 
			\Qcircuit @C=0.5em @R=0.4em {
				&\gate{\rm T} & \gate{\rm T}&\qw & \push{\rule{.1em}{0em}=\rule{.1em}{0em}}&& \gate{\rm S}&\qw &&
			&\gate{\rm H}&\gate{\rm Z}&\gate{\rm H} &\push{\rule{.1em}{0em}=\rule{.1em}{0em}}&& \gate{\rm X}&\qw   &&
              &\gate{\rm H}&\gate{\rm T}&\gate{\rm H} &\push{\rule{.1em}{0em}=\rule{.1em}{0em}}&& \gate{\rm R_x(\frac{\pi}{4})}&\qw 
	}
		};
	\end{tikzpicture}
	\begin{tikzpicture}
		\node at (5.5, 5){ 
			\Qcircuit @C=0.5em @R=0.4em {
	&\gate{\rm H}&\gate{\rm S}&\gate{\rm H} &\qw& \push{\rule{.1em}{0em}=\rule{.1em}{0em}}&&\gate{\rm S^{\dagger}}&\gate{\rm H}&\gate{\rm S^{\dagger}}&\qw 
	&&&\gate{\rm H}&\gate{\rm S^{\dagger}}&\gate{\rm H}&\qw & \push{\rule{.1em}{0em}=\rule{.1em}{0em}}&& \gate{\rm S}&\gate{\rm H}&\gate{\rm S} &\qw 	}
		};
	\end{tikzpicture}
	\begin{tikzpicture}
	\node at (5.5, 5){ 
			\Qcircuit @C=0.5em @R=0.4em {
			&\qw &\qw &\ctrl{2} &\qw &\qw & \qw& && \qw &\ctrl{2} &\qw&\qw   &&&&  \qw &\qw &\ctrl{2} &\qw &\qw & \qw& && \qw &\ctrl{2} &\qw&\qw \\
			&&&&&& &\push{\rule{.1em}{0em}=\rule{.1em}{0em}}&&&& &&   &&&     &&&&&& \push{\rule{.1em}{0em}=\rule{.1em}{0em}}&&& \\
			&\gate{\rm H}&\gate{\rm S}&\targ&\gate{\rm S^{\dagger}}&\gate{\rm H}&\qw& && \gate{\rm S^{\dagger}} &\targ &\gate{\rm S}&\qw && && \gate{\rm H}&\gate{\rm S^{\dagger}}&\targ&\gate{\rm S}&\gate{\rm H}&\qw& && \gate{\rm S} &\targ &\gate{\rm S^{\dagger}}&\qw   \\
			}};
	\end{tikzpicture}
	\begin{tikzpicture}
	\node at (5.5, 5){	\Qcircuit @C=0.5em @R=0.4em {
			 &  \qswap  & \qw &&&&  &  \ctrl{2} &  \targ  &  \ctrl{2}  &\qw &&&&  &  \ctrl{2} &   \gate{\rm H} &\ctrl{2} &\gate{\rm H}     	&\ctrl{2}&    \qw 
			 &&&\gate{\rm H}&\ctrl{2}&\gate{\rm H} &\qw  &&& \targ &\qw \\
			&		\qwx	&&&\push{\rule{.1em}{0em}=\rule{.1em}{0em}}&&&					&			
			&		&      	& 		&	\push{\rule{.1em}{0em}=\rule{.1em}{0em}}					&					&				&					&         			&&&&	&&	
		&&  &&&\push{\rule{.1em}{0em}=\rule{.1em}{0em}} 	\\
			 &   \qswap\qwx	   		&       \qw &&&&     	&   \targ      		&  \ctrl{-2}    &   \targ      		&   \qw   &&&&    &   \targ      		&   \gate{\rm H}      	&   \targ      		&\gate{\rm H} 		&\targ      		&   \qw  
		&&	& \gate{\rm H}&\targ&\gate{\rm H} &\qw  &&& \ctrl{-2}&\qw \\	 
			&			&&&&		&  	&					&					&					&       		& 					&						&					&				&					&         			&&&&			 
		}};
	\end{tikzpicture}
 \begin{tikzpicture}
	\node at(5.5, 5) {
	\Qcircuit @C=0.5em @R=0.4em {
	 &\targ&\qswap &\qw  &&& \ctrl{2}&\targ&\qw &&&
	 &\ctrl{2} &\targ&\qw &&&  \targ &\ctrl{2}&\qw &&&
	 \ctrl{2} &\targ &\ctrl{2}&\qw &&& \targ&\qw &&&
	 \\
	 &&\qwx&& \push{\rule{.1em}{0em}=\rule{.1em}{0em}} &&&  
	 &&&&&&&& \push{\rule{.1em}{0em}=\rule{.1em}{0em}} &&&   
	 &&&&&&& \push{\rule{.1em}{0em}=\rule{.1em}{0em}} &&&   
	 &&&&
	 \\
	 &\ctrl{-2}&\qswap\qwx&\qw &&& \targ &\ctrl{-2}&\qw &&&
	 &\targ &\targ&\qw   &&&  \qw &\targ &\qw &&
	 &\targ &\qw &\targ&\qw &&& \targ&\qw  &&&
	}
	};
	\end{tikzpicture}
	\begin{tikzpicture}
	\node at(5.5, 5) {
	\Qcircuit @C=0.5em @R=0.4em {
	 &\qw &\ctrl{1}&\qw &\qw &&& \ctrl{1}&\ctrl{2}&\qw &&&
	 &\ctrl{1} &\qw &\ctrl{1}&\qw &&& \qw &\ctrl{2}&\qw &&
	 &\qw &\ctrl{2}&\ctrl{1}&\qw &&& \ctrl{1}&\qw&\qw &
	 \\
	 &\ctrl{1} &\targ&\ctrl{1} &\qw&\push{\rule{.1em}{0em}=\rule{.1em}{0em}}&&
	 \targ&\qw &\qw  &&&
	 &\targ&\ctrl{1}&\targ&\qw&\push{\rule{.1em}{0em}=\rule{.1em}{0em}}&&
	 \ctrl{1}&\qw &\qw  &&
	 &\ctrl{1}& \qw &\targ&\qw &\push{\rule{.1em}{0em}=\rule{.1em}{0em}} && \targ &\ctrl{1} &\qw 
	 \\
	 &\targ&\qw &\targ &\qw  &&& \qw &\targ &\qw  &&&
	 &\qw &\targ &\qw &\qw &&& \targ &\targ&\qw &&
	 &\targ &\targ &\qw&\qw  &&&  \qw &\targ &\qw
	}
	};
	\end{tikzpicture}
	\end{center}
	\caption{The 1-qubit  gate and 2-qubit gate cancellation rules.}
	\label{redrule}
\end{figure}

%% file: sections/commutation.tex
\begin{figure}[htb]
    \begin{center}
    \scalefont{1}
    \begin{tikzpicture}
    \node at (5.5, 5){
    \Qcircuit @C=0.5em @R=0.4em {
        & \gate{\rm S}&\gate{\rm T} &\qw & \push{\rule{.1em}{0em}=\rule{.1em}{0em}}&& \gate{\rm T}&\gate{\rm S}&\qw &&
         & \gate{\rm S^{\dagger}}&\gate{\rm T} &\qw & \push{\rule{.1em}{0em}=\rule{.1em}{0em}}&& \gate{\rm T}&\gate{\rm S^{\dagger}}&\qw 
    }
    };
    \end{tikzpicture}
    \begin{tikzpicture}
    \node at (5.5, 5){
    \Qcircuit @C=0.5em @R=0.4em {
          &\gate{\rm T^{\dagger}} & \gate{\rm S}&\qw & \push{\rule{.1em}{0em}=\rule{.1em}{0em}}&& \gate{\rm S}&\gate{\rm T^{\dagger}}&\qw  &&
          &\gate{\rm T^{\dagger}} & \gate{\rm S^{\dagger}}&\qw & \push{\rule{.1em}{0em}=\rule{.1em}{0em}}&& \gate{\rm S^{\dagger}}&\gate{\rm T^{\dagger}}&\qw
    }
    };
    \end{tikzpicture}
    \begin{tikzpicture}
    \node at (5.5, 5){
    \Qcircuit @C=0.5em @R=0.4em{
        &\qw&\qw&\ctrl{2}&\qw&\qw &&& \qw&\ctrl{2}&\qw&\qw&\qw && &\qw&\ctrl{2}&\qw&\ctrl{2}&\qw   &&&&\ctrl{2}&\qw&\ctrl{2}&\qw&\qw  &&
        \\
        &&&&&&\push{\rule{.1em}{0em}=\rule{.1em}{0em}}&&  &&&&&&&  &&&  &&& \push{\rule{.1em}{0em}=\rule{.1em}{0em}} &&  &&&   &&&&&&& &&  &&
        \\
        &\gate{\rm R_z}&\gate{\rm H}&\targ&\gate{\rm H}&\qw && &\gate{\rm H}&\targ&\gate{\rm H}&\gate{\rm R_z}&\qw &&& \gate{\rm R_z}&\targ&\gate{\rm R_z'}&\targ&\qw &&  &&\targ&\gate{\rm R_z'}&\targ&\gate{\rm R_z}&\qw
        &&
    }
    };
    \end{tikzpicture}
    \begin{tikzpicture}
        \node at (5.5, 5){
    \Qcircuit @C=0.5em @R=0.4em{
    &\qw & \ctrl{2}&\qw  &&& \ctrl{2} &\qw &\qw &&
    &\ctrl{2}&\qw&\qw  &&&  \qw &\ctrl{2}&\qw &&
    &\gate{\rm R_z}&\ctrl{2}&\qw &&& \ctrl{2}&\gate{\rm R_z}&\qw 
        \\
      &&& &\push{\rule{.1em}{0em}=\rule{.1em}{0em}} &  &&&  &&
    &\qw &\ctrl{1}&\qw &\push{\rule{.1em}{0em}=\rule{.1em}{0em}}&& \ctrl{1}&\qw &\qw &&
    &&&& \push{\rule{.1em}{0em}=\rule{.1em}{0em}} &&& 
        \\
      &\gate{\rm R_x}& \targ&\qw  &&& \targ &\gate{\rm R_x} &\qw &&
      &\targ &\targ &\qw  &&& \targ &\targ &\qw & &
      &\qw & \targ &\qw &&&\targ &\qw&\qw
    }
    };
    \end{tikzpicture}
    \begin{tikzpicture}
    \node at (5.5, 2){
    \Qcircuit @C=0.5em @R=0.4em {
    &\ctrl{2}&\ctrl{1}&\qw  &&& \ctrl{1}&\ctrl{2}&\qw  &&
    &\ctrl{1} &\qw&\qw&\qw&\qw  &&&  \qw&\qw&\qw&\ctrl{1}&\qw &&
    \\
    &\qw &\targ& \qw &\push{\rule{.1em}{0em}=\rule{.1em}{0em}}&& \targ&\qw &\qw  &&
    &\targ&\gate{\rm H}&\ctrl{1}&\gate{\rm H}&\qw &\push{\rule{.1em}{0em}=\rule{.1em}{0em}}&& \gate{\rm H}&\ctrl{1}&\gate{\rm H}&\targ&\qw   &&
    \\
    &\targ &\qw &\qw  &&& \qw &\targ &\qw  &&
    &\qw&\qw&\targ&\qw&\qw  &&&\qw &\targ&\qw&\qw&\qw &&
    \\
    }
    };
    \end{tikzpicture}
    \end{center}
    \caption{The commutation gate rules.}
    \label{comrule}
\end{figure}

%% file: sections/pattern_surf.tex
\begin{figure}[htb]
	\begin{center}
	\scalefont{1}
	\begin{tikzpicture}
		\node at (5.5, 5){ 
			\Qcircuit @C=0.5em @R=0.4em {
            &\qw&\gate{\rm Z}&\qw 
            &&\push{\rule{.1em}{0em}= \rule{.1em}{0em}}&&  &\qw&\gate{\rm X}&\gate{\rm Y}&\qw
	}
		};
	\end{tikzpicture}
	\\
	\begin{tikzpicture}
		\node at (5.5, 5){ 
			\Qcircuit @C=0.5em @R=0.4em {
            &\qw&\gate{\rm H}&\qw 
            &&\push{\rule{.1em}{0em}= \rule{.1em}{0em}}&&  &\qw&\gate{\rm R_y(-\frac{\pi}{2})}&\gate{\rm Z}&\qw
            &&\push{\rule{.1em}{0em}= \rule{.1em}{0em}}&&  &\qw&\gate{\rm Z}&\gate{\rm R_y(\frac{\pi}{2})}&\qw
            &&\push{\rule{.1em}{0em}= \rule{.1em}{0em}}&&  &\qw&\gate{\rm X}&\gate{\rm R_y(-\frac{\pi}{2})}&\qw
	}
		};
	\end{tikzpicture}
	\\
	\begin{tikzpicture}
		\node at (5.5, 5){ 
			\Qcircuit @C=0.5em @R=0.4em {
            &\qw&\gate{\rm T}&\qw 
            &&\push{\rule{.1em}{0em}= \rule{.1em}{0em}}&&  &\qw&\gate{\rm H}&\gate{\rm R_x(\frac{\pi}{4})}&\gate{\rm H}&\qw
            &&\push{\rule{.1em}{0em}= \rule{.1em}{0em}}&&
            &\qw&\gate{\rm R_y(\frac{\pi}{2})}&\gate{\rm R_x(\frac{\pi}{4})}&\gate{\rm R_y(-\frac{\pi}{2})}&\qw
	}
		};
	\end{tikzpicture}
	\\
	\begin{tikzpicture}
		\node at (5.5, 5){ 
			\Qcircuit @C=0.5em @R=0.4em {
            &\qw&\gate{\rm T^{\dagger}}&\qw 
            &&\push{\rule{.1em}{0em}= \rule{.1em}{0em}}&&  &\qw&\gate{\rm H}&\gate{\rm R_x(-\frac{\pi}{4})}&\gate{\rm H}&\qw
            &&\push{\rule{.1em}{0em}= \rule{.1em}{0em}}&&
            &\qw&\gate{\rm R_y(\frac{\pi}{2})}&\gate{\rm R_x(-\frac{\pi}{4})}&\gate{\rm R_y(-\frac{\pi}{2})}&\qw
	}
		};
	\end{tikzpicture}
	\\
	\begin{tikzpicture}
		\node at (5.5, 5){ 
			\Qcircuit @C=0.5em @R=0.4em {
            &\qw&\gate{\rm S}&\qw 
            &&\push{\rule{.1em}{0em}= \rule{.1em}{0em}}&&  &\qw&\gate{\rm H}&\gate{\rm R_x(\frac{\pi}{2})}&\gate{\rm H}&\qw
            &&\push{\rule{.1em}{0em}= \rule{.1em}{0em}}&&
            &\qw&\gate{\rm R_y(\frac{\pi}{2})}&\gate{\rm R_x(\frac{\pi}{2})}&\gate{\rm R_y(-\frac{\pi}{2})}&\qw
	}
		};
	\end{tikzpicture}
	\\
	\begin{tikzpicture}
		\node at (5.5, 5){ 
			\Qcircuit @C=0.5em @R=0.4em {
            &\qw&\gate{\rm S^{\dagger}}&\qw 
            &&\push{\rule{.1em}{0em}= \rule{.1em}{0em}}&&  &\qw&\gate{\rm H}&\gate{\rm R_x(\frac{\pi}{2})}&\gate{\rm H}&\qw
            &&\push{\rule{.1em}{0em}= \rule{.1em}{0em}}&&
            &\qw&\gate{\rm R_y(\frac{\pi}{2})}&\gate{\rm R_x(-\frac{\pi}{2})}&\gate{\rm R_y(-\frac{\pi}{2})}&\qw
	}
		};
	\end{tikzpicture}
	\\
	\begin{tikzpicture}
		\node at (5.5, 5){ 
			\Qcircuit @C=0.5em @R=0.4em {
            &\ctrl{1}&\qw 
            &\push{\rule{.1em}{0em}= \rule{.1em}{0em}}&&
            &\qw &\ctrl{1}&\qw&\qw
            \\
            &\targ&\qw 
            &&&
            &\gate{\rm R_y(-\frac{\pi}{2})}&\ctrl{-1}&\gate{\rm R_y(\frac{\pi}{2})}&\qw
	}
		};
	\end{tikzpicture}
	\\
	\begin{tikzpicture}
		\node at (5.5, 5){ 
			\Qcircuit @C=0.5em @R=0.4em {
            &\qswap  &\qw 
            &\push{\rule{.1em}{0em}= \rule{.1em}{0em}}&&
            &\qw &\ctrl{1}&\gate{\rm R_y(-\frac{\pi}{2})}&\ctrl{1}&\gate{\rm R_y(\frac{\pi}{2})}&\ctrl{1}&\qw&\qw
            \\
            &\qswap\qwx &\qw 
            &&&
            &\gate{\rm R_y(-\frac{\pi}{2})}&\ctrl{-1}&\gate{\rm R_y(\frac{\pi}{2})}&\ctrl{-1}&\gate{\rm R_y(-\frac{\pi}{2})}&\ctrl{-1}&\gate{\rm R_y(\frac{\pi}{2})}&\qw
	}
		};
	\end{tikzpicture}
	\\
	\begin{tikzpicture}
		\node at (5.5, 5){ 
		\Qcircuit @C=0.5em @R=0.4em {
 		 & \ctrl{2}  & \qw 	&&& \gate{\rm T} &\targ &\qw & \gate{\rm T^{\dagger}} &\targ &\qw & \gate{\rm T}  &\targ & \gate{\rm T^{\dagger}} &\targ &\qw \\
		 & \ctrl{1}  	 & \qw  &\push{\rule{.1em}{0em}= \rule{.1em}{0em}}&&  \gate{\rm T}&\qw&\ctrl{1}&\qw&\ctrl{-1} &\ctrl{1}&\qw&\qw &\qw&\ctrl{-1}&\qw \\	   
		 &\gate{\rm Z}  	 & \qw  &&& \qw &\ctrl{-2}&\targ& \gate{\rm T^{\dagger}} & \qw&\targ&\qw&\ctrl{-2}& \gate{\rm T}&\qw&\qw\\	   			}};
	\end{tikzpicture}
	\end{center}
	\caption{Gate decomposition into primitives supported in the superconducting Surface-17 processor.}
	\label{surf}
\end{figure}
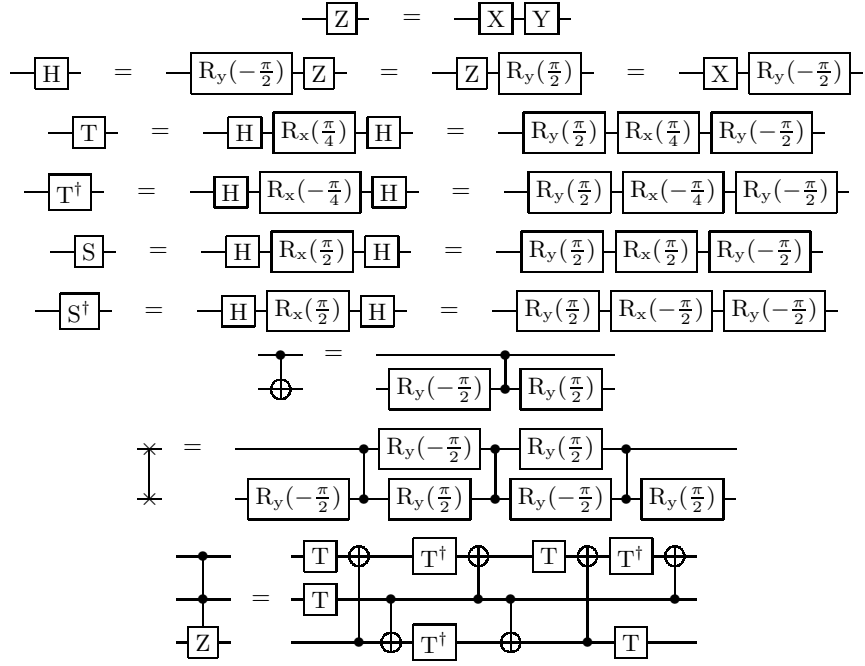

%% file: sections/tof3_after.tex
\newcommand{\TRY}[1]{\rm {R_y(#1)}}
\newcommand{\TRX}[1]{\rm {R_x(#1)}}

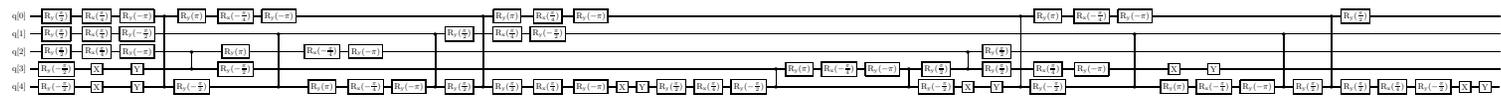
\begin{sidewaysfigure}
    \centering
        \tikzset{global scale/.style={
    scale=#1,
    every node/.append style={scale=#1}
  }
}
    \begin{tikzpicture}[global scale=0.35]
		\node at (5.5, 5){ 
		\Qcircuit @C=1em @R=0.4em @!R{
		  \lstick{\rm q[0]}&\gate{\TRY{\frac{\pi}{2}}} &\gate{\rm \TRX{\frac{\pi}{4}}}&\gate{\rm \TRY{-\pi}}& \ctrl{4}&\gate{\rm \TRY{\pi}}&\gate{\rm \TRX{-\frac{\pi}{4}}} &\gate{ \rm \TRY{-\pi}}&\qw&\qw&\qw&\qw &\qw& \ctrl{4}&\gate{\TRY{\pi}}&\gate{\rm \TRX{\frac{\pi}{4}}}&\gate{\TRY{-\pi}} &\qw &\qw&\qw&\qw&\qw&\qw&\qw&\qw&\qw&\qw&\qw&\qw&\qw&\ctrl{4}&\gate{\TRY{\pi}}&\gate{\rm \TRX{-\frac{\pi}{4}}}&\gate{\TRY{-\pi}}& \qw& \qw&\qw& \qw& \qw&\ctrl{4}&\gate{\TRY{\frac{\pi}{2}}}&\qw&\qw&\qw&\qw&\qw\\
		 \lstick{\rm q[1]}&\gate{\TRY{\frac{\pi}{2}}}  &\gate{\rm \TRX{\frac{\pi}{4}}}&\gate{\TRY{-\frac{\pi}{2}}} & \qw& \qw&\qw  &\ctrl{3}&\qw&\qw&\qw&\ctrl{3}&\gate{\TRY{\frac{\pi}{2}}}&\qw&\gate{\rm \TRX{\frac{\pi}{4}}}&\gate{\TRY{-\frac{\pi}{2}}}&\qw &\qw&\qw&\qw&\qw&\qw&\qw&\qw&\qw&\qw&\qw&\qw&\qw&\qw& \qw& \qw& \qw&\ctrl{3}& \qw& \qw&\qw&\ctrl{3}& \qw& \qw& \qw&\qw&\qw&\qw&\qw&\qw\\
		  \lstick{\rm q[2]}&\gate{\TRY{\frac{\pi}{2}}} &\gate{\rm \TRX{\frac{\pi}{4}}}&\gate{\rm \TRY{-\pi}}& \qw & \ctrl{1}&\gate{\rm \TRY{\pi}}& \qw&\gate{\rm \TRX{-\frac{\pi}{4}}}&\gate{\TRY{-\pi}}&\qw&\qw&\qw&\qw &\qw &\qw &\qw &\qw &\qw&\qw&\qw&\qw&\qw&\qw&\qw&\qw&\qw&\qw&\ctrl{1}&\gate{\TRY{\frac{\pi}{2}}}& \qw& \qw& \qw&\qw& \qw& \qw&\qw& \qw& \qw& \qw& \qw&\qw&\qw&\qw&\qw&\qw\\
		  \lstick{\rm q[3]}&\gate{\TRY{-\frac{\pi}{2}}}&\gate{\rm X} &\gate{\rm Y}&\qw &\ctrl{-1}&\gate{\TRY{-\frac{\pi}{2}}} & \qw&\qw&\qw&\qw&\qw&\qw &\qw &\qw &\qw &\qw &\qw&\qw&\qw&\qw&\qw&\ctrl{1}&\gate{\rm \TRY{\pi}}&\gate{\rm \TRX{-\frac{\pi}{4}}}&\gate{\rm \TRY{-\pi}}&\ctrl{1} &\gate{\TRY{\frac{\pi}{2}}} &\ctrl{-1}&\gate{\TRY{\frac{\pi}{2}}}& \qw&\gate{\rm \TRX{\frac{\pi}{4}}}&\gate{\TRY{-\pi}}&\qw&\gate{\rm X} &\gate{\rm Y}& \qw& \qw& \qw& \qw& \qw&\qw&\qw&\qw&\qw&\qw\\
 		 \lstick{\rm q[4]}&\gate{\TRY{-\frac{\pi}{2}}} &\gate{\rm X} &\gate{\rm Y}&\ctrl{-4} &\gate{\TRY{-\frac{\pi}{2}}}& \qw& \ctrl{-3}&\gate{\TRY{\pi}}&\gate{\rm \TRX{-\frac{\pi}{4}}}&\gate{\TRY{-\pi}}&\ctrl{-3}&\gate{\TRY{\frac{\pi}{2}}}& \ctrl{-4}&\gate{\TRY{\frac{\pi}{2}}}&\gate{\TRX{\frac{\pi}{4}}}&\gate{\TRY{-\pi}}&\gate{\rm X}&\gate{\rm Y}&\gate{\TRY{\frac{\pi}{2}}}&\gate{\rm \TRX{\frac{\pi}{4}}}&\gate{\TRY{-\frac{\pi}{2}}}&\ctrl{-1}&\qw&\qw&\qw&\ctrl{-1} &\gate{\TRY{-\frac{\pi}{2}}}&\gate{\rm X}&\gate{\rm Y}&\ctrl{-4}&\gate{\TRY{-\frac{\pi}{2}}}& \qw&\ctrl{-3}&\gate{\TRY{\pi}}&\gate{\rm \TRX{-\frac{\pi}{4}}}&\gate{\TRY{-\pi}}&\ctrl{-3}&\gate{\TRY{\frac{\pi}{2}}}&\ctrl{-4}&\gate{\TRY{\frac{\pi}{2}}}&\gate{\rm \TRX{\frac{\pi}{4}}}&\gate{\TRY{-\frac{\pi}{2}}}&\gate{\rm X}&\gate{\rm Y}&\qw\\
 		 }};
	\end{tikzpicture}
  
    \caption{The  quantum circuit Toff-NC$_3$  rewritten to Surface-17 quantum processor.}
      \label{fig_sur_after}
\end{sidewaysfigure}

%% file: sections/16qbt.tex
\begin{figure}
	\centering
	\tikzset{global scale/.style={
		scale=#1,
		every node/.append style={scale=#1}
	  }
	}
	\begin{tikzpicture}[global scale=0.65]
	\node at (2,0){(a)};
	\node at (11,0){(b)};
			\node at (5.5, 5){ 
			\Qcircuit @C=1em @R=0.4em @!R{
			\lstick{\rm q[0]}&\gate{\rm X}&\gate{\rm X}&\qw &\qw &\qw &\qw  &\targ&\qw &\qw&\qw &\qw &\qw &\qw &\qw&\qw &&&&&\lstick{\rm q[0]}&\qw&\qw&\qw &\qw&\targ&\qw &\qw&\qw&\qw \\
				\lstick{\rm q[1]}&\gate{\rm X}&\qw &\qw&\qw&\ctrl{6}&\gate{\rm X}   &\qw &\qw &\qw&\qw &\qw &\qw&\qw &\qw &\qw  &&&&& \lstick{\rm q[1]}& \gate{\rm X}&\qw&\ctrl{6}&\gate{\rm X}&\qw&\qw&\qw&\qw&\qw  \\
				\lstick{\rm q[2]}&\targ&\gate{\rm X}&\qw&\qw &\qw &\qw   &\qw &\qw &\qw &\qw &\qw&\ctrl{4}&\qw &\qw &\qw &&&&& \lstick{\rm q[2]}&\targ&\gate{\rm X}&\qw &\qw&\qw&\ctrl{4}&\qw&\qw&\qw \\
				\lstick{\rm q[3]}&\qw&\gate{\rm X}&\qw &\qw&\qw &\ctrl{5}  &\ctrl{-3}  &\gate{\rm X}&\targ &\targ &\qw &\qw &\qw&\qw &\qw &&&&& \lstick{\rm q[3]} &\qw&\gate{\rm X}&\qw &\ctrl{5}&\ctrl{-3}&\qw&\gate{\rm X}&\qw&\qw\\
				\lstick{\rm q[4]}&\qw &\ctrl{8}&\qw&\qw &\qw &\qw &\qw &\ctrl{5} &\qw &\qw &\qw&\qw&\qw  &\qw &\qw &&&&&  \lstick{\rm q[4]}&\qw &\ctrl{8} &\qw &\qw&\qw&\qw&\qw&\qw&\qw\\
				\lstick{\rm q[5]}&\qw&\qw&\ctrl{5} &\gate{\rm X} &\qw &\qw  &\gate{\rm X}&\qw&\qw  &\qw  &\ctrl{5}&\qw&\qw &\qw &\qw  &&&&& \lstick{\rm q[5]}&\qw&\qw&\qw  &\qw&\ctrl{4}&\qw&\qw&\qw&\qw \\
				\lstick{\rm q[6]}&\qw &\qw&\qw&\gate{\rm X} &\qw &\qw  &\gate{\rm X} &\qw&\qw &\qw &\qw  &\targ&\qw&\qw &\qw  &&&&& \lstick{\rm q[6]}&\qw  &\qw &\qw &\qw&\qw&\targ&\qw&\qw&\qw \\
				\lstick{\rm q[7]}&\qw&\qw&\qw &\qw &\targ&\qw   &\qw &\qw &\qw &\qw &\qw &\qw&\qw&\ctrl{6} &\qw &&&&&\lstick{\rm q[7]} &\qw&\qw&\targ&\qw&\qw&\ctrl{6}&\qw&\qw&\qw \\
				\lstick{\rm q[8]}&\qw&\qw&\qw &\gate{\rm X}&\qw  &\targ &\gate{\rm X}&\qw  &\qw &\qw &\qw &\ctrl{3}&\gate{\rm X}  &\qw&\qw  &&&&& \lstick{\rm q[8]}&\qw&\qw&\qw &\targ&\qw&\qw&\ctrl{3}&\gate{\rm X}&\qw \\
				\lstick{\rm q[9]}&\qw&\qw&\qw&\ctrl{5} &\ctrl{5} &\qw  &\qw &\targ &\qw  &\qw &\qw &\qw&\gate{\rm X}&\qw&\qw  &&&&&\lstick{\rm q[9]} & \qw&\qw&\qw &\qw&\targ&\qw&\qw&\gate{\rm X}&\qw \\
				\lstick{\rm q[10]}&\qw&\qw&\targ&\qw&\qw  &\qw  &\qw&\qw  &\qw  &\qw &\targ &\qw&\qw&\qw &\qw &&&&& \lstick{\rm q[10]}&\qw&\qw&\qw &\qw&\qw&\qw&\qw&\qw&\qw \\
				\lstick{\rm q[11]}&\qw&\qw&\gate{\rm X} &\qw &\qw &\gate{\rm X} &\qw &\qw &\qw &\qw  &\qw &\targ&\gate{\rm X}&\qw &\qw &&&&& \lstick{\rm q[11]}&\qw&\qw&\qw &\qw&\qw&\qw&\targ&\gate{\rm X}&\qw \\
				\lstick{\rm q[12]}&\qw&\targ&\qw&\qw &\qw &\qw  &\qw &\qw  &\qw &\qw &\qw &\qw&\qw&\qw &\qw &&&&& \lstick{\rm q[12]} &\qw&\targ&\qw &\qw&\qw&\qw&\qw&\qw&\qw\\
				\lstick{\rm q[13]}&\ctrl{-11}&\gate{\rm X}&\qw &\qw&\qw &\qw   &\qw &\qw&\ctrl{-10}  &\ctrl{-10} &\qw&\qw &\qw &\targ &\qw &&&&&\lstick{\rm q[13]} &\ctrl{-11} &\gate{\rm X}&\qw &\qw&\qw&\targ&\qw&\qw&\qw \\
				\lstick{\rm q[14]}&\qw&\qw &\qw&\targ&\targ &\gate{\rm X}  &\qw &\qw&\qw  &\qw &\qw&\qw &\qw&\qw &\qw &&&&&\lstick{\rm q[14]} &\gate{\rm X}&\qw&\qw &\qw&\qw&\qw&\qw&\qw&\qw \\
				\lstick{\rm q[15]}&\gate{\rm X}&\gate{\rm X}&\qw &\qw &\qw&\qw  &\qw&\qw &\qw  &\qw &\qw &\qw &\qw&\qw &\qw &&&&&\lstick{\rm q[15]} &\qw&\qw&\qw &\qw&\qw&\qw&\qw&\qw&\qw \gategroup{10}{5}{15}{6}{.7em}{--}  \gategroup{4}{10}{14}{11}{.7em}{--}
				\\
			}};
\end{tikzpicture}
	\caption{(a) The  quantum circuit 16QBT\_05YCTFL\_3. (b) The  quantum circuit 16QBT\_05YCTFL\_3  optimized by QSRewriting. }
	\label{fig_16qbt}
  \end{figure}
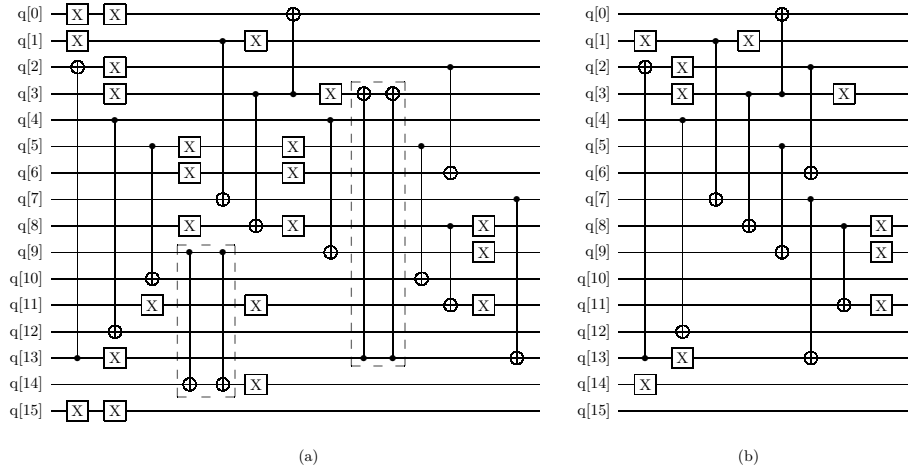

%% file: sections/exp1.tex
    \definecolor{cyan1}{RGB}{0,255,255}
    \definecolor{red1}{RGB}{255,20,147}
  \pgfplotsset{every axis/.append style={
font=\small,
}
}
\begin{figure*}[htb]
  \centering
  \subfigure[]{ 

  \begin{tikzpicture}
 \scalefont{0.8}
  \begin{axis}[
    legend style={
at={(0.5,0.9)},legend columns=2,legend cell align=left,
anchor=center},
 enlargelimits=0,
      width=11cm,
      xtick align=outside,
      ytick align=outside,
      xmin=-50,xmax=2200,ymin=0,ymax=700,
      xtick={0,600,...,2200},
      xlabel={(d$_1$, d$_2$)},
      ylabel={1-qubit gate count},
       xticklabels={(0.0, 0.1), (0.1, 0.3),  (0.2, 0.6), (0.5, 0.1), (0.7, 0.1)},
      no markers
      ]
      \addplot[color=cyan1,ybar interval=0] table [x=x,y=b] {./sections/spin_single.dat};
      \addplot+[color=yellow] table [color=yellow,x=x,y=c] {./sections/spin_single.dat};
      \addplot+[color=red1] table [color=red,x=x,y=d] {./sections/spin_single.dat};
      \addplot+[color=blue] table [color=blue,x=x,y=e] {./sections/spin_single.dat};
      \legend{1-qubit\_gate\_count,QSRewriting,QGRewriting,PaF}
    \end{axis}
  \end{tikzpicture}}
  \subfigure[]{
  \begin{tikzpicture}
 \scalefont{0.8}
  \begin{axis}[
    legend style={
at={(0.5,0.9)},legend columns=2,legend cell align=left,
anchor=center},
      enlargelimits=0.005,
      width=11cm,
      xtick align=outside,
      ytick align=outside,
      xlabel={(d$_1$, d$_2$)},
      ylabel={2-qubit gate count},
      xmin=-50,xmax=2200,ymin=0,ymax=450,
      xtick={0,600,...,2200},
      xticklabels={(0.0, 0.1), (0.1, 0.3),  (0.2, 0.6), (0.5, 0.1), (0.7, 0.1)},
      bar width=2pt,
      no markers
      ]
      \addplot[color=cyan1,ybar interval=0] table [x=x,y=b] {./sections/spin_two.dat};
      \addplot+[color=yellow] table [color=yellow,x=x,y=c] {./sections/spin_two.dat};
      \addplot+[color=red1] table [color=red,x=x,y=d] {./sections/spin_two.dat};
      \addplot+[color=blue] table [color=blue,x=x,y=e] {./sections/spin_two.dat};
      \legend{2-qubit\_gate\_count,QSRewriting,QGRewriting,PaF}    \end{axis}
  \end{tikzpicture}}
  \end{figure*} 
  \begin{figure*}[htb]
  \centering
  \ContinuedFloat
    \subfigure[]{
  \begin{tikzpicture}
 \scalefont{0.8}
    \begin{axis}[
    legend style={
at={(0.5,0.9)},legend columns=2,legend cell align=left,
anchor=center},
      enlargelimits=0.005,
      width=11cm,
      xtick align=outside,
      ytick align=outside,
      xmin=-50,xmax=2200, ymin=0, ymax=700,
      xtick={0,600,...,2200},
      xlabel={(d$_1$, d$_2$)},
      ylabel={total gate count},
      xticklabels={(0.0, 0.1), (0.1, 0.3),  (0.2, 0.6), (0.5, 0.1), (0.7, 0.1)},
      bar width=0.1pt,
      no markers
      ]
      \addplot[color=cyan1,ybar interval=0] table [x=x,y=b] {./sections/spin_num.dat};
      \addplot+[color=yellow] table [color=yellow,x=x,y=c] {./sections/spin_num.dat};
      \addplot+[color=red1] table [color=red,x=x,y=d] {./sections/spin_num.dat};
      \addplot+[color=blue] table [color=blue,x=x,y=e] {./sections/spin_num.dat};
      \legend{total\_gate\_count,QSRewriting,QGRewriting,PaF};   
    \end{axis}
  \end{tikzpicture}}
  \subfigure[]{
   \begin{tikzpicture}
 \scalefont{0.8}
      \begin{axis}[
    legend style={
at={(0.5,0.9)},legend columns=2,legend cell align=left,
anchor=center},
      enlargelimits=0.005,
      width=11cm,
      xtick align=outside,
      ytick align=outside,
      xmin=-50,xmax=2200,ymin=0,ymax=65,
      xtick={0,600,...,2200},
      xlabel={(d$_1$, d$_2$)},
      ylabel={depth},
      xticklabels={(0.0, 0.1), (0.1, 0.3),  (0.2, 0.6), (0.5, 0.1), (0.7, 0.1)},
      bar width=1pt,
      no markers
      ]
      \addplot[color=cyan1,ybar interval=0] table [x=x,y=b] {./sections/spin_depth.dat};
      \addplot+[color=yellow] table [color=yellow,x=x,y=c] {./sections/spin_depth.dat};
      \addplot+[color=red1] table [color=red,x=x,y=d] {./sections/spin_depth.dat};
      \addplot+[color=blue] table [color=blue,x=x,y=e] {./sections/spin_depth.dat};
      \legend{depth,QSRewriting,QGRewriting,PaF}   
    \end{axis}
  \end{tikzpicture}}
\caption{Comparison of the 1-qubit gate, 2-qubit gate, total gate counts, and depths of the  quantum circuits generated by QSRewriting, QGRewriting and PaF.}
    \label{spin_qrewriting}
\end{figure*}
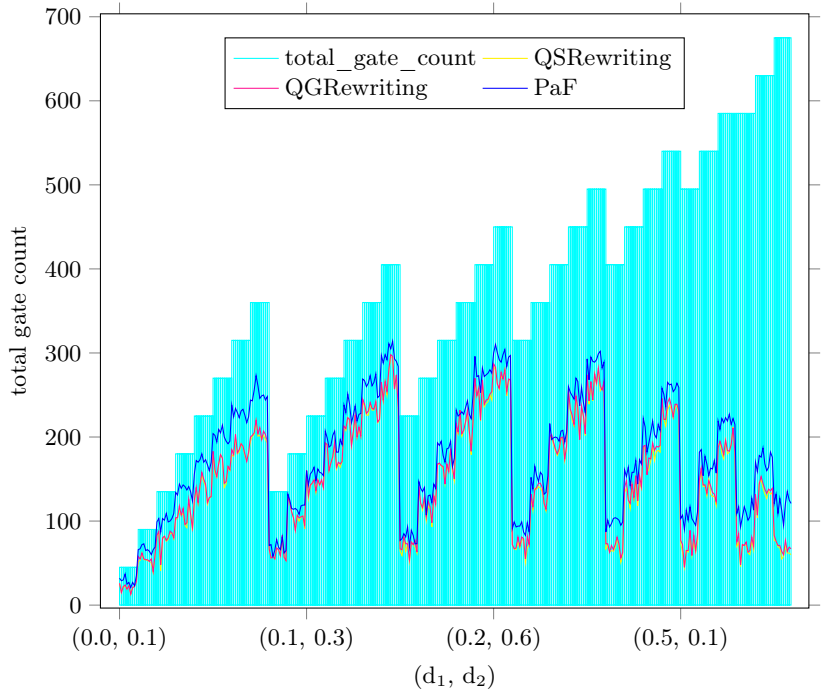
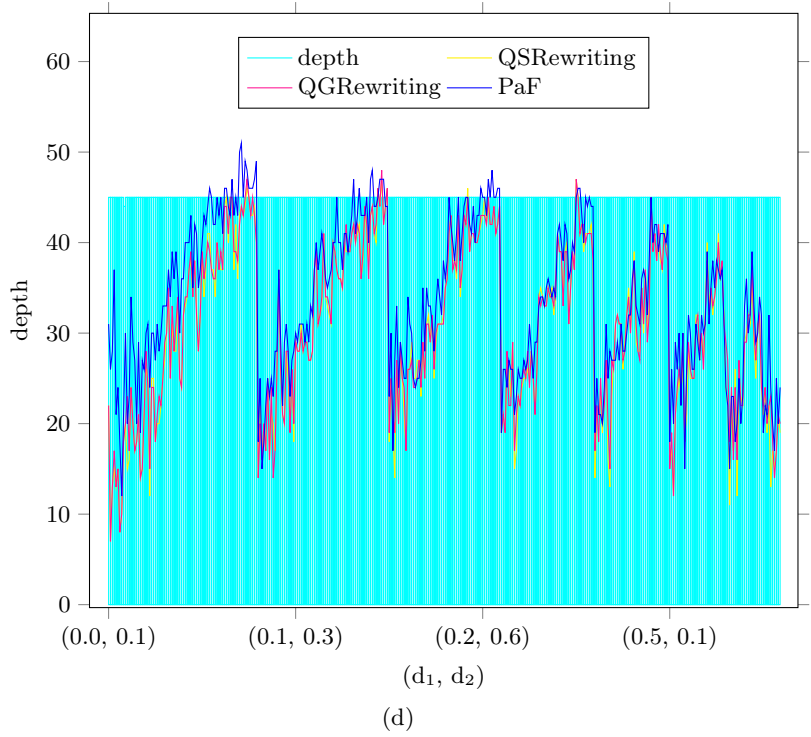 

%% file: sections/tabs.tex
\begin{table*}[htb]\tiny
			\centering
			\caption{\label{tof_table_num}Comparison of the gate counts of circuits.}\vspace{1mm}
			\footnotesize{
				\begin{tabular*}{\linewidth}{p{2.5cm}<{\centering}|p{0.6cm}<{\centering}p{1cm}<{\centering}p{0.9cm}<{\centering}p{0.8cm}<{\centering}p{1.2cm}<{\centering}p{0.9cm}<{\centering}p{1.2cm}<{\centering}p{1cm}<{\centering}p{1.3cm}<{\centering}}
					\hline		benchmark&$n$&$g$&$g_0$&$t_0$&$g_1$&$t_1$&$g_2$&$t_2$ &$\Delta$\\
					\hline
Toff-NC$_{3}$&5&9&45&0.00 &135&0.03 &80&0.50&40.74\% \\
Toff-Barenco$_{3}$&5&10&58&0.00 &174&0.07 &101&0.52&41.95\% \\
Mod 5$_{4}$&5&15&63&0.00 &187&0.09 &89&0.57&52.41\% \\
Toff-NC$_{4}$&7&15&75&0.00 &225&0.10 &134&0.89&40.44\% \\
Toff-Barenco$_{4}$&7&18&114&0.00 &342&0.19 &198&1.12&42.11\% \\
Toff-NC$_{5}$&9&21&105&0.00 &315&0.18 &188&1.34&40.32\% \\
Toff-Barenco$_{5}$&9&26&170&0.01 &510&0.32 &296&1.70&41.96\% \\
VBE-Adder$_{3}$&10&30&150&0.00 &450&0.34 &266&1.51&40.89\% \\
GF($2^{4}$)-Mult&12&33&225&0.01 &675&0.58 &388&2.68&42.52\% \\
Mod-Mult$_{55}$&9&35&119&0.00 &341&0.20 &211&1.04&38.12\% \\
GF($2^{5}$)-Mult&15&47&347&0.01 &1041&0.80 &601&3.74&42.27\% \\
CSLA-MUX$_{3}$&15&50&170&0.01 &510&0.44 &315&2.23&38.24\% \\
Toff-NC$_{10}$&19&51&255&0.01 &765&0.49 &458&3.31&40.13\% \\
GF($2^{6}$)-Mult&18&63&495&0.02 &1485&1.19 &854&5.36&42.49\% \\
Toff-Barenco$_{10}$&19&66&450&0.01 &1350&0.95 &786&4.76&41.78\% \\
RC-Adder$_{6}$&14&68&200&0.01 &584&0.93 &361&2.66&38.18\% \\
Mod-Red$_{21}$&11&74&278&0.01 &786&0.93 &463&3.39&41.09\% \\
GF($2^{7}$)-Mult &21&81&669&0.02 &2007&1.61 &1153&7.37&42.55\% \\
CSUM-MUX$_{9}$&30&84&420&0.01 &1204&1.57 &721&3.98&40.12\% \\
QCLA-Com$_{7}$&24&95&443&0.01 &1299&1.02 &778&5.98&40.11\% \\
QCLA-Adder$_{10}$&36&113&521&0.01 &1563&1.31 &957&7.29&38.77\% \\
GF($2^{8}$)-Mult&24&115&883&0.02 &2649&2.35 &1516&12.95&42.77\% \\
GF($2^{9}$)-Mult&27&123&1095&0.03 &3285&2.72 &1885&12.09&42.62\% \\
GF($2^{10}$)-Mult&30&147&1347&0.03 &4041&3.36 &2316&14.90&42.69\% \\
QCLA-Mod$_{7}$&26&176&884&0.02 &2638&2.06 &1570&12.24&40.49\% \\
Adder$_{8}$&24&216&900&0.02 &2676&6.79 &1623&13.00&39.35\% \\
GF($2^{16}$)-Mult&48&363&3435&0.13 &10305&9.44 &5865&38.96&43.09\% \\
Mod-Adder$_{1024}$&28&865&4285&0.09 &12855&18.19 &7403&57.77&42.41\% \\
GF($2^{32}$)-Mult&96&1305&13593&0.30 &40779&38.71 &23069&157.25&43.43\% \\
GF($2^{64}$)-Mult&192&4539&53691&1.17 &161073&146.63 &91065&635.20&43.46\% \\
GF($2^{128}$)-Mult&384&17275&213883&5.59 &641649&584.97 &362429&3656.48&43.52\% \\
GF($2^{131}$)-Mult&393&18333&224265&5.90 &672795&616.90 &379766&3927.99&43.55\% \\
GF($2^{163}$)-Mult&489&27705&346533&9.67 &1039599&945.51 &587034&7452.82&43.53\% \\
\hline
				\end{tabular*} \vspace*{.2mm}
				\\\vspace{1mm}\parbox{12cm}{
					Note: $n$: the number of qubits.
                    $g$: the gate count of  the target circuit.
					$g_0$: the  gate count of the target circuit after decomposition without internal
					optimization on $G_{IBM}$ gate set.
                    $g_1$: the gate count of the target circuit without internal optimization on $G_{Sur}$ gate set.
					$g_2$: the gate count of the target circuit with internal optimization on $G_{Sur}$ gate set.
					$t_0-t_2 $: running time in seconds.
                    $\Delta$: $(g_1-g_2)/g_1 \times 100\%$.
				}
			}
		\end{table*}	
\begin{table*}[htb]
\tiny
			\centering
			\caption{\label{tof_table_dep}Comparison of the depths of circuits. }\vspace{1mm}
			\footnotesize{
				\begin{tabular*}{\linewidth}{p{2.5cm}<{\centering}|p{0.6cm}<{\centering}p{1cm}<{\centering}p{0.9cm}<{\centering}p{0.9cm}<{\centering}p{1.2cm}<{\centering}p{1cm}<{\centering}p{1cm}<{\centering}p{1cm}<{\centering}p{1.3cm}<{\centering}}
					\hline		benchmark&$n$&$d$&$d_0$&$t_0$&$d_1$&$t_1$&$d_2$&$t_2$ &$\Delta$\\
					\hline
Toff-NC$_{3}$&5&7&23&0.00 &64&0.03 &42&0.50&34.38\% \\
Toff-Barenco$_{3}$&5&9&31&0.00 &86&0.07 &52&0.52&39.53\% \\
Mod 5$_{4}$&5&15&36&0.00 &97&0.09 &49&0.57&49.48\% \\
Toff-NC$_{4}$&7&11&38&0.00 &104&0.10 &67&0.89&35.58\% \\
Toff-Barenco$_{4}$&7&17&61&0.00 &166&0.19 &102&1.12&38.55\% \\
Toff-NC$_{5}$&9&15&53&0.00 &144&0.18 &92&1.34&36.11\% \\
Toff-Barenco$_{5}$&9&25&91&0.01 &246&0.32 &152&1.70&38.21\% \\
VBE-Adder$_{3}$&10&20&70&0.00 &194&0.34 &113&1.51&41.75\% \\
GF($2^{4}$)-Mult&12&17&85&0.01 &236&0.58 &145&2.68&38.56\% \\
Mod-Mult$_{55}$&9&14&43&0.00 &118&0.20 &80&1.04&32.20\% \\
GF($2^{5}$)-Mult&15&20&111&0.01 &310&0.80 &187&3.74&39.68\% \\
CSLA-MUX$_{3}$&15&17&59&0.01 &166&0.44 &107&2.23&35.54\% \\
Toff-NC$_{10}$&19&35&128&0.01 &344&0.49 &217&3.31&36.92\% \\
GF($2^{6}$)-Mult&18&25&139&0.02 &384&1.19 &235&5.36&38.80\% \\
Toff-Barenco$_{10}$&19&65&241&0.01 &646&0.95 &402&4.76&37.77\% \\
RC-Adder$_{6}$&14&28&93&0.01 &261&0.93 &166&2.66&36.40\% \\
Mod-Red$_{21}$&11&43&141&0.01 &383&0.93 &238&3.39&37.86\% \\
GF($2^{7}$)-Mult &21&29&166&0.02 &458&1.61 &280&7.37&38.86\% \\
CSUM-MUX$_{9}$&30&15&53&0.01 &147&1.57 &96&3.98&34.69\% \\
QCLA-Com$_{7}$&24&15&70&0.01 &192&1.02 &115&5.98&40.10\% \\
QCLA-Adder$_{10}$&36&15&64&0.01 &182&1.31 &111&7.29&39.01\% \\
GF($2^{8}$)-Mult&24&39&199&0.02 &544&2.35 &335&12.95&38.42\% \\
GF($2^{9}$)-Mult&27&36&219&0.03 &606&2.72 &367&12.09&39.44\% \\
GF($2^{10}$)-Mult&30&40&246&0.03 &680&3.36 &412&14.90&39.41\% \\
QCLA-Mod$_{7}$&26&39&172&0.02 &487&2.06 &284&12.24&41.68\% \\
Adder$_{8}$&24&55&191&0.02 &527&6.79 &315&13.00&40.23\% \\
GF($2^{16}$)-Mult&48&71&415&0.13 &1136&9.44 &699&38.96&38.47\% \\
Mod-Adder$_{1024}$&28&521&2218&0.09 &6397&18.19 &3775&57.77&40.99\% \\
GF($2^{32}$)-Mult&96&137&849&0.30 &2324&38.71 &1447&157.25&37.74\% \\
GF($2^{64}$)-Mult&192&263&1711&1.17 &4688&146.63 &2856&635.20&39.08\% \\
GF($2^{128}$)-Mult&384&517&3437&5.59 &9420&584.97 &5750&3656.48&38.96\% \\
GF($2^{131}$)-Mult&393&537&3526&5.90 &9658&616.90 &5902&3927.99&38.89\% \\
GF($2^{163}$)-Mult&489&665&4390&9.67 &12026&945.51 &7310&7452.82&39.22\% \\
					\hline
				\end{tabular*} \vspace*{.2mm}
				\\\vspace{1mm}\parbox{12cm}{
					Note: $n$: the number of qubits.
                    $d$: the depth of the target circuit.
                    $d_0$: the depth of the target circuit after decomposition without internal optimization on $G_{IBM}$ gate set.
                    $d_1$: the depth of  the rewritten circuit without internal optimization on $G_{Sur}$ gate set.
                    $d_2$: the depth of  the rewritten circuit with internal optimization on $G_{Sur}$ gate set.
					$t_0-t_2 $: running time in seconds.
                    $\Delta$: $(d_1-d_2)/d_1\times 100\%$. 

				}
			}
		\end{table*}

%% file: pattern.bbl
\begin{thebibliography}{10}
\providecommand{\url}[1]{\texttt{#1}}
\providecommand{\urlprefix}{URL }
\providecommand{\doi}[1]{https://doi.org/#1}

\bibitem{Abdessaied13}
Abdessaied, N., Soeken, M., Wille, R., Drechsler, R.: Exact template matching
  using boolean satisfiability. In: 2013 IEEE 43rd International Symposium on
  Multiple-Valued Logic. pp. 328--333 (2013)

\bibitem{Amy2018}
Amy, M., Azimzadeh, P., Mosca, M.: On the controlled-{NOT} complexity of
  controlled-{NOT}{\textendash}phase circuits. Quantum Science and Technology
  \textbf{4}(1),  015002 (2018)

\bibitem{Arute19}
Arute, F., Arya, K., et~al.: Quantum supremacy using a programmable
  superconducting processor. Nature  \textbf{574},  505–510 (2019)

\bibitem{IBM127}
Ball, P.: First quantum computer to pack 100 qubits enters crowded race. Nature
   \textbf{599}, ~542 (2021)

\bibitem{DP2003}
Bellman, R.E.: Dynamic Programming. Dover Publications, Inc., USA (2003)

\bibitem{chen2021}
Chen, M., Zhang, Y., Li, Y.: A quantum circuit optimization framework based on
  pattern matching. SPIN  \textbf{11}(03),  2140008 (2021)

\bibitem{Dumitrescu_2018}
Dumitrescu, E., McCaskey, A., Hagen, G., Jansen, G., Morris, T., Papenbrock,
  T., Pooser, R., Dean, D., Lougovski, P.: Cloud quantum computing of an atomic
  nucleus. Physical Review Letters  \textbf{120}(21) (may 2018)

\bibitem{Gro96}
Grover, L.K.: A fast quantum mechanical algorithm for database search. In:
  Proc. 28th Annual {ACM} Symposium on the Theory of Computing. pp. 212--219.
  {ACM} (1996)

\bibitem{HHL09}
Harrow, A.W., Hassidim, A., Lloyd, S.: Quantum algorithm for solving linear
  systems of equations. Physical Review Letters  \textbf{103}(15),  150502
  (2009)

\bibitem{qiskit}
IBM: Qiskit: An open-source {SDK} for working with quantum computers at the
  level of pulses, circuits, and algorithms. \url{https://github.com/QISKit}
  (2020)

\bibitem{Raban20}
Iten, R., Moyard, R., Metger, T., Sutter, D., Woerner, S.: Exact and practical
  pattern matching for quantum circuit optimization (2020)

\bibitem{Iwama2002}
Iwama, K., Kambayashi, Y., Yamashita, S.: Transformation rules for designing
  cnot-based quantum circuits. In: Proceedings of the 39th Annual Design
  Automation Conference. p. 419–424. Association for Computing Machinery
  (2002)

\bibitem{Zhihao2019}
Jia, Z., Padon, O., Thomas, J., Warszawski, T., Zaharia, M., Aiken, A.: Taso:
  Optimizing deep learning computation with automatic generation of graph
  substitutions. In: Proceedings of the 27th ACM Symposium on Operating Systems
  Principles. p. 47–62. Association for Computing Machinery, New York, NY,
  USA (2019)

\bibitem{Kissinger2020}
Kissinger, A., van~de Wetering, J.: {PyZX}: Large scale automated diagrammatic
  reasoning. Electronic Proceedings in Theoretical Computer Science
  \textbf{318},  229--241 (may 2020)

\bibitem{Kjaergaard2020}
Kjaergaard, M., Schwartz, M.E., Braumüller, J., Krantz, P., Wang, J.I.J.,
  Gustavsson, S., Oliver, W.D.: Superconducting qubits: Current state of play.
  Annual Review of Condensed Matter Physics  \textbf{11}(1),  369--395 (2020)

\bibitem{Kjaergaard20}
Kjaergaard, M., Schwartz, M.E., Braumüller, J., Krantz, P., Wang, J.I.J.,
  Gustavsson, S., Oliver, W.D.: Superconducting qubits: Current state of play.
  Annual Review of Condensed Matter Physics  \textbf{11}(1),  369–395 (2020)

\bibitem{LaoSAA22}
Lao, L., van Someren, H., Ashraf, I., Almud{\'{e}}ver, C.G.: Timing and
  resource-aware mapping of quantum circuits to superconducting processors.
  {IEEE} Trans. Comput. Aided Des. Integr. Circuits Syst.  \textbf{41}(2),
  359--371 (2022)

\bibitem{Liu21}
Liu, J., Bello, L., Zhou, H.: Relaxed peephole optimization: A novel compiler
  optimization for quantum circuits. In: 2021 IEEE/ACM International Symposium
  on Code Generation and Optimization. pp. 301--314 (2021)

\bibitem{Mad22}
Madsen, L.S., Laudenbach, F., Askarani, M.F., Rortais, F., Vincent, T., Bulmer,
  J.F.F., Miatto, F.M., Neuhaus, L., Helt, L.G., Collins, M.J., Lita, A.E.,
  Gerrits, T., Nam, S.W., Vaidya, V.D., Menotti, M., Dhand, I., Vernon, Z.,
  Quesada, N., Lavoie, J.: Quantum computational advantage with a programmable
  photonic processor. Nature  \textbf{606},  75--81 (2022)

\bibitem{McKeeman65}
McKeeman, W.M.: Peephole optimization. Commun. ACM  \textbf{8}(7),  443–444
  (1965)

\bibitem{Prakash2019full}
Murali, P., Linke, N.M., Martonosi, M., Abhari, A.J., Nguyen, N.H., Alderete,
  C.H.: Full-stack, real-system quantum computer studies: Architectural
  comparisons and design insights. In: 2019 ACM/IEEE 46th Annual International
  Symposium on Computer Architecture (ISCA). pp. 527--540 (2019)

\bibitem{YSN2017}
Nam, Y.S., Ross, N.J., Su, Y., Childs, A.M., Maslov, D.: Automated optimization
  of large quantum circuits with continuous parameters. npj Quantum Information
   \textbf{4}, ~23 (2018)

\bibitem{Nielsen2016}
Nielsen, M.A., Chuang, I.L.: Quantum Computation and Quantum Information (10th
  Anniversary edition). Cambridge University Press (2016)

\bibitem{Pointing2021}
Pointing, J., Padon, O., Jia, Z., Ma, H., Hirth, A., Palsberg, J., Aiken, A.:
  Quanto: Optimizing quantum circuits with automatic generation of circuit
  identities (2021)

\bibitem{Aditya06}
Prasad, A.K., Shende, V.V., Markov, I.L., Hayes, J.P., Patel, K.N.: Data
  structures and algorithms for simplifying reversible circuits. {ACM} J.
  Emerg. Technol. Comput. Syst.  \textbf{2}(4),  277--293 (2006)

\bibitem{Preskill2018}
Preskill, J.: Quantum {C}omputing in the {NISQ} era and beyond. {Quantum}
  \textbf{2}, ~79 (Aug 2018)

\bibitem{Rahman12}
Rahman, M.M., Dueck, G.W.: Optimal quantum circuits of three qubits. In: 2012
  IEEE 42nd International Symposium on Multiple-Valued Logic. pp. 161--166
  (2012)

\bibitem{Sho94}
Shor, P.W.: Algorithms for quantum computation: Discrete logarithms and
  factoring. In: Proc. 35th Annual Symposium on Foundations of Computer
  Science. pp. 124--134. {IEEE} Computer Society (1994)

\bibitem{Sivarajah2020}
Sivarajah, S., Dilkes, S., Cowtan, A., Simmons, W., Edgington, A., Duncan, R.:
  t|ket>: a retargetable compiler for {NISQ} devices. Quantum Science and
  Technology  \textbf{6}(1),  014003 (nov 2020)

\bibitem{markskilbeck}
Skilbeck, M., Peterson, E., appleby, Davis, E., Karalekas, P., Bello-Rivas,
  J.M., Kochmanski, D., Beane, Z., Smith, R., Shi, A., Scott, C., Paszke, A.,
  Hulburd, E., Young, M., Jackson, A.S., BHAVISHYA, Alam, M.S.,
  Velázquez-Rodríguez, W., c.~b. osborn, fengdlm, jmackeyrigetti:
  rigetti/quilc: v1.21.0 (Jul 2020)

\bibitem{Soeken16}
Soeken, M., Dueck, G.W., Rahman, M.M., Miller, D.M.: An extension of
  transformation-based reversible and quantum circuit synthesis. In: 2016 IEEE
  International Symposium on Circuits and Systems (ISCAS). pp. 2290--2293
  (2016)

\bibitem{Tan2020}
Tan, B., Cong, J.: Optimality study of existing quantum computing layout
  synthesis tools. IEEE Transactions on Computers  (Jul 2020)

\bibitem{xu2022quartz}
Xu, M., Li, Z., Padon, O., Lin, S., Pointing, J., Hirth, A., Ma, H., Palsberg,
  J., Aiken, A., Acar, U.A., Jia, Z.: Quartz: Superoptimization of quantum
  circuits. In: Proceedings of the 43rd ACM SIGPLAN Conference on Programming
  Language Design and Implementation (2022)

\bibitem{2019Zhang}
Zhang, Y., Deng, H., Li, Q., Song, H., Nie, L.: Optimizing quantum programs
  against decoherence: Delaying qubits into quantum superposition. In: 2019
  International Symposium on Theoretical Aspects of Software Engineering. pp.
  184--191. IEEE (2019)

\end{thebibliography}
